\documentclass{jpsj3}
\usepackage{txfonts}

\newcommand{\be}{\begin{equation}}
\newcommand{\ee}{\end{equation}}
\newcommand{\eq}[1]{Eq.~(\ref{#1})}
\newcommand{\fig}[1]{Fig.~\ref{#1}}
\def\bea{\begin{eqnarray}}
\def\eea{\end{eqnarray}}

\def\bra{\langle}
\def\ket{\rangle}

\def\vk{{\bf k}}

\def\vr{{\bf r}}

\title{Molecular Anisotropy and Electron-Momentum Distributions from Randomly Oriented Molecules}

\author{Hiroyuki Yamase\thanks{yamase.hiroyuki@nims.go.jp}}
\inst{Research Center for Materials Nanoarchitectonics (MANA), National Institute for Materials Science (NIMS), Tsukuba 305-0047, Japan}

\abst{Compton scattering provides direct access to electron momentum distributions, but quantitative interpretation of measurements from molecular ensembles requires proper treatment of molecular orientations. In particular, although an individual molecule generally has an anisotropic electron momentum distribution, molecules in gases and isotropic liquids are randomly oriented. Here we establish a direct connection between the Compton profile of a molecule with a fixed orientation and one observed from an isotropic molecular ensemble. Starting from the molecular Kohn--Sham equation, we derive the orientationally averaged Compton profile and obtain explicit analytical expressions for its evaluation using the 6-31G($d$,$p$) basis set. Applying the formalism to H$_{2}$, N$_{2}$, CO$_{2}$, O$_{2}$, H$_{2}$O, NH$_{3}$, CH$_{4}$, and C$_{2}$H$_{5}$OH, we find that the effect of molecular orientation is strongly molecule dependent. Most notably, ethanol exhibits an almost isotropic Compton profile despite its strongly anisotropic molecular structure. This result demonstrates that real-space molecular anisotropy does not, in general, determine the degree of anisotropy of the Compton profile. Our results establish a quantitative framework for interpreting Compton-scattering measurements from randomly oriented molecular systems and provide a microscopic basis for relating molecular structure to experimentally accessible Compton profile. 
}

\begin{document}
\maketitle

\section{Introduction}
Compton scattering provides a one-dimensional projection of the electron momentum distribution, which is called Compton profile, and has long been an important tool for investigating electronic structure in condensed matter \cite{cooper85, cooper}. With the development of high-brilliance synchrotron sources, Compton-scattering techniques have also been extended to molecular and disordered systems, including gases \cite{sakurai16,sakurai21,mizuno26,yamase26a}, liquids \cite{manninen77,hakala06,ono16}, and combustion environments \cite{mizuno26,yamase26a}. These applications open the possibility of probing electronic structure in systems, for which conventional momentum-space measurements are difficult or impossible.

A fundamental difference between molecular and crystalline systems is the role of molecular orientation. In a crystal, the relative orientation of the electronic structure with respect to the experimental geometry is fixed, whereas molecules in a gas or an isotropic liquid are generally distributed over all spatial orientations. Consequently, the Compton profile measured from a molecular ensemble is not simply the profile of a molecule in a particular orientation, but an average over molecular orientations. A quantitative interpretation of molecular Compton-scattering measurements, therefore, requires a well-defined connection between the Compton profile of an individual molecule and that of a randomly oriented ensemble.

More importantly, the effect of molecular orientation is not determined solely by the geometrical anisotropy of the molecule. An individual molecule can have a strongly nonspherical structure in real space, while its Compton profile may be substantially more isotropic. Conversely, relatively simple molecular structures can produce pronounced directional dependence in momentum space. This raises a basic question: to what extent is molecular anisotropy in real space reflected in the anisotropy of Compton profile?  Establishing this relationship is important not only for interpreting molecular Compton profiles, but also for understanding what information about molecular electronic structure can be extracted from momentum-space measurements.

Theoretical calculations of molecular Compton profiles have a long history. Isotropic Compton profiles have been evaluated using quantum-chemical wave functions and Gaussian basis sets \cite{saenz97}, and general computational approaches have been developed for arbitrary Gaussian basis functions \cite{lehtola11}. These studies provide efficient and accurate methods for obtaining isotropic Compton profiles. However, the present formulation provides a different method that allows the orientation dependence itself to be quantified. In fact, an explicit formulation that retains the molecular orientation before performing the orientational average provides an additional advantage: it allows the orientation dependence of the Compton profile to be calculated and quantified directly. This distinction is particularly useful for addressing the relation between molecular structure in real space and Compton profile in momentum space.

In this work, we formulate the Compton profile of a randomly oriented molecular ensemble by explicitly averaging the fixed-orientation molecular Compton profile over the full solid angle. Starting from the molecular Kohn--Sham equation, we derive the electron momentum distribution and the corresponding Compton profile for an arbitrary molecular orientation. We then perform the orientational average appropriate to an isotropic ensemble and derive explicit analytical expressions for practical calculations using the 6-31G($d$,$p$)  Gaussian basis set. The resulting formulation provides a direct route from molecular orbitals obtained in standard quantum-chemical calculations to the Compton profile expected for a randomly oriented molecular ensemble.

We apply the formalism to H$_{2}$, N$_{2}$, CO$_{2}$, O$_{2}$, H$_{2}$O, NH$_{3}$, CH$_{4}$, and C$_{2}$H$_{5}$OH. We find that the influence of molecular orientation on the Compton profile varies strongly among molecules. In particular, ethanol exhibits the almost isotropic Compton profile despite its strongly nonspherical molecular structure. This result demonstrates that geometrical anisotropy in real space does not directly translate into the anisotropy of Compton profile. The comparison among different molecules therefore provides a quantitative view of how molecular structure and electronic wave functions determine the anisotropy of Compton profile. 

These results establish an explicit theoretical framework for interpreting Compton-scattering measurements from randomly oriented molecular systems and, more broadly, clarify the nontrivial relation between molecular structure in real space and experimentally accessible electron momentum distributions.

The present paper is organized as following.  In Sect.~2, starting from the molecular Kohn--Sham equation, we derive the momentum distribution for a fixed molecular orientation and subsequently perform the orientational average appropriate to an isotropic molecular ensemble. In Sect.~3, we further derive explicit working expressions for the widely used 6-31G($d$,$p$) basis set, providing a practical implementation of the formalism. Equations are very lengthly and explicit expressions are given in Appendices A--C. In Sect.~4, by applying the formalism to H$_{2}$, N$_{2}$, CO$_{2}$, O$_{2}$, H$_{2}$O, NH$_{3}$, CH$_{4}$, and C$_{2}$H$_{5}$OH, we show that the magnitude of the orientational effect is strongly molecule dependent.

\section{Theoretical formalism} 
In this section, we derive a theoretical expression for the orientationally averaged Compton profile relevant to gases and isotropic liquids. We first introduce the electron momentum distribution and the corresponding Compton profile for a fixed molecular orientation. We then formulate the spherical average over all molecular orientations. 

\subsection{Electron momentum distribution and Compton profile}
We consider molecular orbitals obtained from the Kohn-Sham equation,  
\begin{equation}
H\Psi_m(\mathbf r)=E_m\Psi_m(\mathbf r),
\label{kohn-sham}
\end{equation}
where $\Psi_m(\mathbf r)$ and $E_m$ denote the molecular orbital $m$ and its eigenenergy, respectively.
The momentum-space wave function is defined by
\be
\varphi_{m}(\vk) = \int \frac{d \vk}{(\sqrt{2 \pi})^{3}} \Psi_{m}(\vr) {\rm e}^{- i\vk \cdot \vr}  \,.
\ee
The electron momentum density is then given by
\be
n(\vk) = 2 \sum_{m}f(E_{m})\varphi_{m}^{*}(\vk) \varphi_{m}(\vk) \,,
\label{nkT}
\ee
where 
\be
f(E_{m}) = \frac{1}{{\rm e}^{(E_{m}-\mu)/T}+1}
\ee
with temperature $T$ and the chemical potential $\mu$; the factor 2 counts the spin degrees of freedom. Note that $\mu$ is not necessarily zero---we need to determine it to reproduce the number of electrons in a molecule. At zero temperature, Eq.~(\ref{nkT}) reduces to the usual expression obtained by summing over occupied molecular orbitals only.

The Compton profile is obtained by projecting the electron momentum density onto the momentum-transfer direction of incident photons, conventionally chosen as the $k_z$ axis,
\be
J(k_{z}) =  \int d k_{x}  d k_{y}  n(\vk)  \,.
\label{profile}
\ee
Substituting Eq.~(\ref{nkT}) into Eq.~(\ref{profile}) yields
\be
J(k_{z}) = 2 \sum_{m} f(E_{m}) \int d k_{x}  d k_{y}  \varphi_{m}^{*}(\vk)\varphi_{m}(\vk) \,.
\label{profile2}
\ee

\subsection{Orientational average}
Equation~(\ref{profile2}) describes the Compton profile for a fixed molecular orientation. In gases or isotropic   liquids, however, the molecular orientations are distributed without a preferred direction. The experimentally observed Compton profile, therefore, corresponds to an orientational average over all molecular configurations.

To describe this average, we introduce a rotated coordinate system $(k_1,k_2,k_3)$ related to $(k_x,k_y,k_z)$ through
 \be
\begin{pmatrix}
k_{x} \\ k_{y}\\ k_{z}
\end{pmatrix} 
=
\begin{pmatrix}
\cos \theta \cos \phi & -\sin \phi & \sin \theta \cos \phi \\
\cos \theta \sin \phi & \cos \phi & \sin \theta \sin \phi \\
-\sin \theta & 0 & \cos \theta
\end{pmatrix} 
\begin{pmatrix}
k_{1} \\
k_{2} \\
k_{3}
\end{pmatrix} \,. 
\label{kxk1}
\ee
The Compton profile for a given orientation is
\be
J(k_{3}; \theta, \phi)= 2 \sum_{m}f(E_{m}) \int d k_{1} d k_{2}   | \varphi_{m}(k_{1}, k_{2}, k_{3}; \theta, \phi) |^{2} \,.
\label{Jk3}  
\ee
The orientationally averaged Compton profile is then obtained as 
\be
\bra J(k_{3})  \ket = \frac{1}{4\pi} \int_{0}^{\pi} d \theta \int_{0}^{2\pi}d \phi  J(k_{3}; \theta, \phi) \sin \theta \,.
\label{Jk3-2}
\ee
Equation~(\ref{Jk3-2}) gives the Compton profile appropriate to an isotropic ensemble of randomly oriented molecules.

\section{Implementation using the 6-31G(\boldmath{$d$},\boldmath{$p$}) basis set}
Equation~(\ref{Jk3-2}) provides the expression for the spherically averaged Compton profile. Its practical evaluation, however, requires the momentum-space molecular orbitals. Since quantum-chemistry packages usually provide molecular orbitals in real space, analytical Fourier transformation of the basis functions is necessary.

The formalism is not intrinsically restricted to 6-31G($d$,$p$); this basis is used because it provides a practical and widely adopted test case for an explicit analytical implementation. Hence as a practical implementation we present explicit working formulae for the 6-31G$(d,p)$ basis set. The key idea is that all required momentum-space integrals can be reduced to a finite set of elementary Gaussian integrals, allowing efficient evaluation of the Compton profile.

Here we have in mind molecules composed of atoms in the first and second rows of the periodic table. The wavefunction may be given by 
\be
\Psi_{m}(\vr)= \sum_{i} \sum_{\nu} c(m, i, \nu) g_{\nu}(\vr - {\bf R}_{i}) \,,
\label{wave0}
\ee
where $m$ denotes a molecular orbital; $i$ specifies the atomic position ${\bf R}_{i}$; $\nu$ is the basis function specified by 1$s$6, 2$s$3, 2$p_{x}$3, 2$p_{y}$3, 2$p_{z}$3, 2$s$1, 2$p_{x}$1, 2$p_{y}$1, 2$p_{z}$1, 3$d_{xx}$, 3$d_{yy}$, 3$d_{zz}$, 3$d_{xy}$, 3$d_{xz}$, 3$d_{yz}$ for non-hydrogen molecules, and $\nu$=1$s$3, 2$s$1, 2$p_{x}$1, 2$p_{y}$1, 2$p_{z}$1 for hydrogen; see Appendix A for details. 
The basis function $g_{\nu}(\vr)$ is decomposed into 
\be
g_{\nu}(\vr)= \sum_{a=1}^{a(\nu)} b_{\nu}^{a} g_{\nu}^{0} (\vr, \lambda_{\nu}^{a})\,,
\label{g0}
\ee
where $g_{\nu}^{0}$ is Gaussian primitive function.

When we do Fourier transformation of \eq{wave0}, we obtain 
\be
\varphi_{m}(\vk) = \sum_{i} \sum_{\nu} c(m, i, \nu) \sum_{a=1}^{a(\nu)} b_{\nu}^{a} f_{\nu}^{0} (\vk, \lambda_{\nu}^{a}) {\rm e}^{-i\vk \cdot {\bf R}_{i}}\,.
\ee
Explicit forms of $g^{0}_{\nu}$ and $f^{0}_{\nu}$ are given in Appendix~A. 
We have to compute 
\be
| \varphi_{m}(k_{x}, k_{y}, k_{z}) |^{2} = \sum_{i_{1}}\sum_{i_{2}}\sum_{\nu_{1}}\sum_{\nu_{2}}\sum_{a_{1}}\sum_{a_{2}} c(m, i_{1}, \nu_{1}) c(m, i_{2}, \nu_{2}) b_{\nu_{1}}^{a_{1}} b_{\nu_{2}}^{a_{2}} f_{\nu_{1}}^{0}(\vk, \lambda_{\nu_{1}}^{a_{1}})f_{\nu_{2}}^{0*}(\vk, \lambda_{\nu_{2}}^{a_{2}}) {\rm e}^{-i \vk \cdot \boldsymbol{\tau}}\,,
\label{varphi2}
\ee
where {\boldmath{$\tau$}}= ${\bf R}_{i_{1}} - {\bf R}_{i_{2}}$. $c(m, i, \nu)$ and $b_{\nu}^{a}$ are real. 
Although Eq.~(\ref{varphi2}) contains complex phase factors arising from the Fourier transformation, the final expression is manifestly real after combining each term with its complex-conjugate counterpart.

Using \eq{kxk1}, $(k_{x}, k_{y}, k_{z})$ and $(\tau_{x}, \tau_{y}, \tau_{z})$ are written in terms of $(k_{1}, k_{2}, k_{3})$, $(\tau_{1}, \tau_{2}, \tau_{3})$, $\theta$, and $\phi$.  Since $\varphi_{m}$ consists of Gaussian function, $k_{1}$ and $k_{2}$ integrals are possible analytically. Hence \eq{Jk3} can be written 
\be
J(k_{3}; \theta, \phi) =2 \sum_{m} f(E_{m})\sum_{i_{1}}\sum_{i_{2}}\sum_{\nu_{1}}\sum_{\nu_{2}}\sum_{a_{1}}\sum_{a_{2}} c(m, i_{1}, \nu_{1}) c(m, i_{2}, \nu_{2})b_{\nu_{1}}^{a_{1}} b_{\nu_{2}}^{a_{2}} I^{a_{1} a_{2}}_{\nu_{1}  \nu_{2}}(k_{3}, \boldsymbol{\tau}; \theta, \phi) \,.
\label{Jk3-3}
\ee
Here $c(m,i,\nu)$ and $b_{\nu}^{a}$ are obtained from a quantum chemistry package and we need to compute 
\be		
I^{a_{1} a_{2}}_{\nu_{1}  \nu_{2}}(k_{3}, \boldsymbol{\tau}; \theta, \phi) =\int d k_{1} d k_{2}{\rm e}^{-ik_{3}\tau_{3}} f_{\nu_{1}}^{0}(\vk, \lambda_{\nu_{1}}^{a_{1}})f_{\nu_{2}}^{0*}(\vk, \lambda_{\nu_{2}}^{a_{2}}) {\rm e}^{-ik_{1}\tau_{1}} {\rm e}^{-ik_{2}\tau_{2}}  \,.
\label{I-1}
\ee
To compute \eq{I-1} efficiently, we recall the explicit form of $f_{\nu}^{0}(\vk, \lambda)$ given in Appendix~A and then consider the following function: 
\be
\widetilde{gg}_{mn}(\lambda, \lambda^{'}, \tau_1, \tau_2) =\int d k_{1} d k_{2} k_{1}^{m}k_{2}^{n} 
{\rm e}^{-ik_{1} \tau_{1}} {\rm e}^{-ik_{2} \tau_{2}} {\rm e}^{-\gamma (k_{1}^{2}+ k_{2}^{2})} \,,
\label{ggmn}
\ee
where $0 \leq m+n \leq 4$ with $m$ and $n$ being integer and $\gamma = \frac{1}{4 \lambda} +\frac{1}{4 \lambda^{'}} = \frac{\lambda + \lambda^{'}}{4 \lambda \lambda^{'}}$. This integral is easily performed. For example, for $m=n=0$ we obtain 
\be
\widetilde{gg}_{00}(\lambda, \lambda', \tau_1, \tau_2) = \frac{\pi}{\gamma} {\rm e}^{-\tau_{1}^{2}/4\gamma} {\rm e}^{-\tau_{2}^{2}/4\gamma} 
\ee
and for $m=1, n=0$
\be
\widetilde{gg}_{10}(\lambda, \lambda', \tau_1, \tau_2) =  \frac{\pi}{\gamma}\frac{-i\tau_{1}}{2\gamma} {\rm e}^{-\tau_{1}^{2}/4\gamma} {\rm e}^{-\tau_{2}^{2}/4\gamma} 
\ee
All remaining integrals up to fourth order can be obtained in the same manner and are summarized in Appendix B.
Since all $\widetilde{gg}_{mn}$ have the same factor ${\rm  e}^{-\tau_{1}^{2}/4\gamma} {\rm e}^{-\tau_{2}^{2}/4\gamma}$, we define $gg_{mn}$ as 
\bea
&&\widetilde{gg}_{mn}=gg_{mn} {\rm  e}^{-\tau_{1}^{2}/4\gamma} {\rm e}^{-\tau_{2}^{2}/4\gamma} \hspace{5mm} {\rm for} \; m+n = \; {\rm even}\,, \\
&&\hspace{10mm} =igg_{mn} {\rm  e}^{-\tau_{1}^{2}/4\gamma} {\rm e}^{-\tau_{2}^{2}/4\gamma} \hspace{4mm} {\rm for} \; m+n = \; {\rm odd}\,. 
\eea

After analytical integration over $k_1$ and $k_2$, all contributions can be reduced to the compact form
\be
I^{a_{1} a_{2}}_{\nu_{1}  \nu_{2}}(k_{3}, \boldsymbol{\tau}; \theta, \phi) = A(\nu_{1},a_{1}) B(k_{3}, \nu_{1},a_{1},\nu_{2},a_{2},\tau_{1}, \tau_{2},\theta,\phi) {\rm e}^{-i k_{3}\tau_{3}}{\rm e}^{-\gamma_{\nu_{1} \nu_{2}}^{a_{1} a_{2}} k_{3}^{2}} {\rm e}^{-(\tau_{1}^{2} +\tau_{2}^{2})/4\gamma_{\nu_{1} \nu_{2}}^{a_{1} a_{2}}}  \,,
\label{I-C}
\ee
where $\gamma_{\nu_{1} \nu_{2}}^{a_{1}a_{2}}=\frac{1}{4\lambda_{\nu_{1}}^{a_{1}}}+ \frac{1}{4\lambda_{\nu_{2}}^{a_{2}}}$. This expression serves as the fundamental building block for evaluating the orientationally averaged Compton profile. 
Equation~(\ref{I-C}) is the key result for practical implementation because all basis-function combinations can be reduced to this form. Functional forms of $A(\nu_{1},a_{1})$ and $B(k_{3}, \nu_{1},a_{1},\nu_{2},a_{2},\tau_{1}, \tau_{2}, \theta, \phi)$ are given in Appendix~C.

\section{Spherically averaged Compton profiles} 
The formalism developed in Sect.~2 is applicable at arbitrary temperature. In the present work, we restrict ourselves to zero temperature because the gap between the highest occupied molecular orbital (HOMO) and the lowest unoccupied molecular orbital (LUMO) in typical molecules corresponds to an energy scale far exceeding room temperature, rendering thermal excitations negligible. Here, the zero-temperature approximation refers only to the electronic occupations. Thermal rotational and vibrational distributions are not included in the present calculation. The effect of these nuclear degrees of freedom on the Compton profile is beyond the scope of the present work. 

Electronic-structure calculations were performed using the GAMESS package \cite{barca20} with the B3LYP hybrid exchange-correlation functional. Initial molecular geometries were taken from the Experimental dataset of the Computational Chemistry Comparison and Benchmark Database (CCC-BDB), and then the  full structural optimization was performed. The resulting molecular orbitals were then transformed into momentum space and used to evaluate the spherically averaged Compton profiles via Eqs.~(\ref{Jk3-2}), (\ref{Jk3-3}), and (\ref{I-C}). For O$_{2}$, the open-shell triplet state is treated explicitly by evaluating the spin-up and spin-down contributions separately.

\begin{figure}[h]
\centering
\includegraphics[width=14cm]{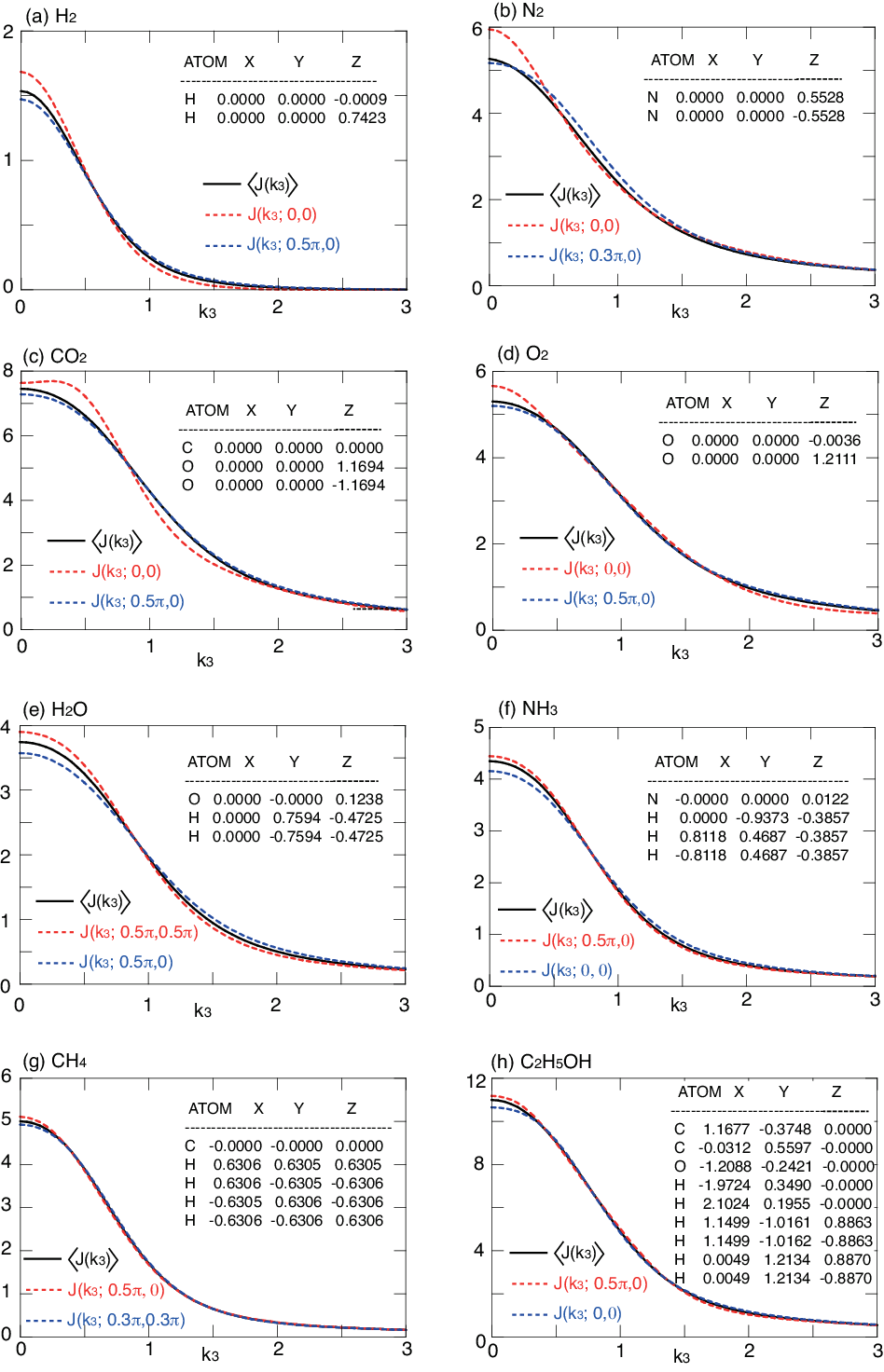}
\caption{(Color online) Compton profiles for (a) H$_{2}$, (b) N$_{2}$, (c) CO$_{2}$, (d) O$_{2}$, (e) H$_{2}$O, (f) NH$_{3}$, (g) CH$_{4}$, and (h) C$_{2}$H$_{5}$OH. The orientationally averaged profile (solid curve) is compared with one calculated in $0 \leq \theta \leq \pi$ and $0 \leq \phi \leq 2 \pi$ with a difference $\Delta \theta = \Delta \phi = 0.1 \pi$. The dashed curves indicate the largest and smallest profiles among the sampled orientations. The optimized molecular geometry is shown in the inset in each panel. 
}
\label{profiles}
\end{figure}

Figure~\ref{profiles} presents the calculated Compton profiles for eight representative molecules: H$_2$, N$_2$, CO$_2$, O$_2$, H$_2$O, NH$_{3}$, CH$_{4}$, and C$_2$H$_5$OH. For comparison, we also show the largest and smallest non-averaged profiles in $0 \leq \theta \leq \pi$ and $0 \leq \phi \leq 2\pi$ with a difference $\Delta \theta = \Delta \phi =0.1\pi$. The optimized molecular geometries are displayed in the insets.

\begin{figure}[h]
\centering
\includegraphics[width=8cm]{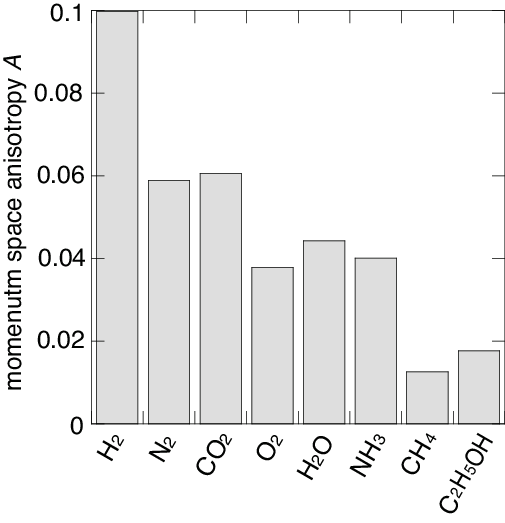}
\caption{Anisotropy of momentum distribution $A$, which is evaluated for the orientation giving the largest deviation from $\bra J(k_{3}) \ket$ in \fig{profiles}. 
}
\label{profiles3}
\end{figure}

To compare more closely the orientationally averaged profile with one calculated for fixed molecular orientation and to quantify its difference, we introduce the dimensionless orientational deviation parameter $A$ defined by 
\be
A=\frac{\int d k_{3} \left|  J(k_{3}; \theta, \phi)  -  \bra J(k_{3}) \ket \right|} {\int d k_{3} \bra J(k_{3})  \ket} \,
\label{A}
\ee
where $\theta$ and $\phi$ is chosen to maximize $A$ in \fig{profiles}. Figure~\ref{profiles3} summarizes the quantity $A$.  A pronounced orientational dependence is observed for H$_2$, N$_2$, CO$_2$, O$_2$, H$_2$O, and NH$_{3}$, whereas the deviation is substantially reduced for CH$_{4}$ and C$_2$H$_5$OH. This systematic variation demonstrates that orientational averaging is not merely a formal requirement, but can quantitatively affect the interpretation of molecular Compton-scattering data.

The weak orientational dependence of ethanol is particularly noteworthy. Although ethanol has a strongly nonspherical molecular geometry, its Compton profile is nearly isotropic over most of the momentum range [\fig{profiles}(h)]. This behavior might be counterintuitive: Given a simple anisotropic wave function, a larger spatial extent along one direction generally corresponds to a narrower momentum distribution along that direction. However, this intuitive correspondence does not necessarily hold for a polyatomic molecule with a complex molecular-orbital structure.

In fact, the electron momentum density is not determined by the Fourier transform of the electron density in real space. Instead it is determined by the squared Fourier transforms of all occupied molecular orbitals [see Eqs.~(\ref{kohn-sham})--(\ref{nkT})].  In ethanol, these orbitals involve multiple atomic centers and different bonding environments, including C-C, C-O, and O-H bonds, as well as electronic contributions localized predominantly around individual atoms. Hence the resulting momentum distribution reflects the combined contributions of orbitals with different spatial characteristics and directional dependences. Such a combination can yield a substantially weaker orientational dependence than what would be expected from the nonspherical molecular geometry alone.

In comparison, H$_{2}$O and NH$_{3}$ exhibit more pronounced orientational dependence in the present calculations. Their valence electronic structures involve directional bonding and lone-pair orbitals associated with a smaller number of atomic centers. These differences in local electronic structure may contribute to the contrast with ethanol, although the present calculations do not establish the contribution of each individual orbital. A quantitative identification of the microscopic origin would require an orbital-resolved analysis of the momentum density. 

The above comparison highlights that the anisotropy of Compton profile is not determined solely by the spatial shape of the molecule. Rather, it reflects the electronic wave functions and their combined contributions in momentum space. Moreover, because the Compton profile is obtained by integrating the momentum density over the two momentum components  [Eqs.~(\ref{profile}) and (\ref{profile2})], the orientational dependence of the profile need not directly reproduce the anisotropy of the full three-dimensional momentum density.

On the other hand, methane exhibits a nearly isotropic momentum distribution. This may be easily understood from its  molecular structure in real space---methane is tetrahedral---it is rather closer to a spherical shape compared with the other molecules we considered. 

It is insightful to point out that the degree of deviation from $\bra J(k_{3})\ket$ is not correlated with the point-group symmetry of molecule. H$_{2}$, N$_{2}$, CO$_{2}$ and O$_{2}$ are $D_{\infty h}$ symmetry; H$_{2}$O and NH$_{3}$ are $C_{2v}$ and $C_{3v}$, respectively; and CH$_{4}$ is $T_{d}$; C$_2$H$_5$OH is $C_{1}$. It is also insightful that the geometry of molecule is not correlated with the parameter $A$. H$_{2}$, N$_{2}$, CO$_{2}$ and O$_{2}$ form a linear structure; H$_{2}$O a bending one; NH$_{3}$ and CH$_{4}$ trigonal pyramidal and tetrahedral ones, respectively; C$_2$H$_5$OH is a strongly nonspherical one.

The present results have a direct implication for the interpretation of synchrotron Compton-scattering experiments on molecular systems. In gases or isotropic liquids, the measured profile should be compared with an orientationally averaged molecular calculation rather than one with the profile of an arbitrarily chosen molecular orientation. The present results further indicate that the magnitude of this correction is molecule dependent. For relatively simple molecules, neglecting orientational averaging can lead to appreciable deviations, whereas for larger molecules such as ethanol the correction may be substantially smaller. This provides a practical criterion for assessing the level of theoretical treatment required when constructing molecular Compton-profile databases or interpreting changes in measured profiles during chemical reactions.

It should be emphasized that the present set of molecules is not intended to establish a universal scaling of the orientational effect with molecular size or complexity. Rather, the calculations demonstrate that the magnitude of the orientational effect is highly molecule dependent and that molecular geometry alone is insufficient to anticipate the degree of anisotropy of Compton profile, 

\section{Conclusions}
We have developed a theoretical framework for describing Compton profiles of randomly oriented molecular systems by explicitly averaging the Compton profile of a fixed-orientation molecule over molecular orientations. Starting from the molecular Kohn--Sham equation, we derived the corresponding electron momentum distribution and Compton profile and obtained explicit analytical expressions for practical calculations using the 6-31G($d$,$p$) Gaussian basis set.

Application to H$_{2}$, N$_{2}$, CO$_{2}$, O$_{2}$, H$_{2}$O, NH$_{3}$, CH$_{4}$, and C$_{2}$H$_{5}$OH demonstrates that the effect of molecular orientation is strongly molecule dependent. The most notable result is obtained for ethanol: despite its strongly nonspherical molecular structure, its Compton profile is nearly isotropic. In contrast, several simpler molecules retain substantial orientation dependence.

This comparison reveals a nontrivial distinction between molecular anisotropy in real space and anisotropy of  Compton profile in momentum space. In particular, the geometrical anisotropy of a molecule alone is not sufficient to predict the anisotropy of the Compton profile. The latter is determined by the momentum-space structure of the occupied molecular orbitals and their contributions to the total electron momentum density.

The present formulation provides more than a procedure for obtaining isotropic molecular Compton profiles: it offers a quantitative framework for examining how molecular structure and electronic wave functions are encoded in momentum space. It can be directly applied to Compton-scattering measurements of gases, isotropic liquids, and other randomly oriented molecular systems, including complex environments such as combustion.

\begin{acknowledgment}
The author thanks T. Oguchi, H. Sakurai, Y. Sakurai, H. Suzuki, and N. Tsuji for telling him the importance of Compton profile. He also thanks A. Nakata for technical aspects to compute the Compton profile.  This work was supported by the Acquisition, Technology \& Logistic Agency (ATLA) (Project No. JPJ004596). He also acknowledges support from the World Premier International Research Center Initiative (WPI), MEXT, Japan.
\end{acknowledgment}

 \appendix
%
%
%

\section{Gaussian functions}
We consider up to 3$d$ electron states in the 6-31G($d,p$) basis set. 
The 6-31G($d$,$p$) basis set is a split-valence Gaussian basis set augmented with polarization functions on both heavy atoms and hydrogen atoms. “6-31G” denotes the valence double-zeta representation, where core orbitals are described by six Gaussian primitives---hence we denote them as $1s6$---and valence orbitals are split into two parts, i.e., three and one primitive Gaussian functions---they are denoted as $2s3$ and $2s1$ for $2s$ orbitals, and $2p_{x(y,z)}3$ and $2p_{x(y,z)}1$ for $2p$ orbital. The “($d$,$p$)” indicates the addition of $d$-type polarization functions on non-hydrogen atoms and $p$-type polarization functions on hydrogen atoms, allowing the ability of the basis to describe anisotropic electron distributions, lone pairs, bent bonds, and molecular response properties. 6-31G($d$,$p$) is commonly employed in density functional theory (DFT) calculations for obtaining reliable molecular structures, vibrational frequencies, and reaction energetics at moderate computational cost. 

Here we provide the original Gaussian function for each orbital. For $s$-orbital, namely $\nu=1s6, 2s3,  2s1$
\be
g_{\nu}^{0}(\vr, \lambda)= \left( \frac{8 \lambda^{3}}{\pi^{3}} \right)^{\frac{1}{4}} {\rm e}^{-\lambda r^{2}}\,,
\ee
where $r=|\vr|=\sqrt{x^{2}+y^{2}+z^{2}}$. 

For $p_{x}$-orbitals, namely $\nu=2p_{x}3, 2p_{x}1$ 
\be
g^{0}_{\nu}(\vr, \lambda) = \left( \frac{128 \lambda^{5}}{\pi^{3}} \right)^{\frac{1}{4}} x {\rm e}^{-\lambda r^{2}} \,.
\ee
Similarly for $p_{y}$ ($\nu=2p_{y}3, 2p_{y}1$) and $p_{z}$ ($\nu=2p_{z}3, 2p_{z}1$) orbitals are, respectively,  
\bea
&&g^{0}_{\nu}(\vr, \lambda) = \left( \frac{128 \lambda^{5}}{\pi^{3}} \right)^{\frac{1}{4}} y {\rm e}^{-\lambda r^{2}} \,,\\
&&g^{0}_{\nu}(\vr, \lambda) = \left( \frac{128 \lambda^{5}}{\pi^{3}} \right)^{\frac{1}{4}} z {\rm e}^{-\lambda r^{2}} \,.
\eea
For a $d$-orbitals, there are five independent orbitals. However, the following six orbitals are considered in the 6-31G($d,p$) basis set: 
\be
g^{0}_{3d_{xx}}(\vr, \lambda) = \left( \frac{2048 \lambda^{7}}{9\pi^{3}} \right)^{\frac{1}{4}} xx {\rm e}^{-\lambda r^{2}} \,,\\
\ee
\be
g^{0}_{3d_{yy}}(\vr, \lambda) = \left( \frac{2048 \lambda^{7}}{9\pi^{3}} \right)^{\frac{1}{4}} yy {\rm e}^{-\lambda r^{2}} \,,\\
\ee
\be
g^{0}_{3d_{zz}}(\vr, \lambda) = \left( \frac{2048 \lambda^{7}}{9\pi^{3}} \right)^{\frac{1}{4}} zz {\rm e}^{-\lambda r^{2}} \,, \\
\ee
\be
g^{0}_{3d_{xy}}(\vr, \lambda) = \left( \frac{2048 \lambda^{7}}{\pi^{3}} \right)^{\frac{1}{4}} xy {\rm e}^{-\lambda r^{2}} \,,\\
\ee
\be
g^{0}_{3d_{xz}}(\vr, \lambda) = \left( \frac{2048 \lambda^{7}}{\pi^{3}} \right)^{\frac{1}{4}} xz {\rm e}^{-\lambda r^{2}} \,, \\
\ee
\be
g^{0}_{3d_{yz}}(\vr, \lambda) = \left( \frac{2048 \lambda^{7}}{\pi^{3}} \right)^{\frac{1}{4}} yz {\rm e}^{-\lambda r^{2}} \,. 
\ee

Their Fourier transform is defined as 
\be
f_{\nu}^{0}(\vk, \lambda)=
\frac{1}{(\sqrt{2\pi})^{3}} \int d x d y d z
g^{0}_{\nu}(\vr, \lambda) {\rm e}^{-i\vk \cdot \vr} \,.
\ee
The calculation of $f_{\nu}^{0}(\lambda, \vk)$ is straightforward. For $\nu=1s6, 2s3$, and 2$s$1, namely for the  $s$-orbital 
\be
f_{\nu}^{0}(\vk, \lambda) = w_{s}(\lambda) {\rm e}^{-\vk^{2}/4\lambda} \,.
\ee
For $\nu=2p_{x}3, 2p_{x}1$
\be 
f_{\nu}^{0}(\vk, \lambda) = iw_{p}(\lambda) k_{x} {\rm e}^{-\vk^{2}/4\lambda} \,.
\ee
For $\nu=2p_{y}3, 2p_{y}1$
\be
f_{\nu}^{0}(\vk, \lambda) = iw_{p}(\lambda) k_{y} {\rm e}^{-\vk^{2}/4\lambda} \,.
\ee
For $\nu=2p_{z}3, 2p_{z}1$
\be
f_{\nu}^{0}(\vk, \lambda) = iw_{p}(\lambda) k_{z} {\rm e}^{-\vk^{2}/4\lambda} \,.
\ee
For 3$d$ orbitals, we obtain
\bea
&&f_{3d_{xx}}^{0}(\vk, \lambda) = w_{dd}(\lambda) (2 \lambda - k_{x}^{2}) {\rm e}^{-\vk^{2}/4\lambda} \,,\\
&&f_{3d_{yy}}^{0}(\vk, \lambda) = w_{dd}(\lambda) (2 \lambda - k_{y}^{2}) {\rm e}^{-\vk^{2}/4\lambda} \,, \\
&&f_{3d_{zz}}^{0}(\vk, \lambda) = w_{dd}(\lambda) (2 \lambda - k_{z}^{2}) {\rm e}^{-\vk^{2}/4\lambda} \,,\\
&&f_{3d_{xy}}^{0}(\vk, \lambda) = - w_{d}(\lambda) k_{x} k_{y} {\rm e}^{-\vk^{2}/4\lambda} \,,\\
&&f_{3d_{xz}}^{0}(\vk, \lambda) = - w_{d}(\lambda) k_{x} k_{z} {\rm e}^{-\vk^{2}/4\lambda} \,,\\
&&f_{3d_{yz}}^{0}(\vk, \lambda) = - w_{d}(\lambda) k_{y} k_{z} {\rm e}^{-\vk^{2}/4\lambda} \,.
\eea
Here 
\bea
&&w_{s}(\lambda) =\left( \frac{8 \lambda^{3}}{\pi^{3}} \right)^{\frac{1}{4}} \frac{1}{(2 \lambda)^{3/2}} \,,\\
&&w_{p}(\lambda) =-\left(\frac{128 \lambda^{5}}{\pi^{3}} \right)^{\frac{1}{4}} \frac{1}{(2 \lambda)^{5/2}}\,, \\
&&w_{dd}(\lambda) =\left( \frac{2048 \lambda^{7}}{9\pi^{3}} \right)^{\frac{1}{4}} \frac{1}{(2 \lambda)^{7/2}} \,,\\
&&w_{d}(\lambda) =\left( \frac{2048 \lambda^{7}}{\pi^{3}} \right)^{\frac{1}{4}} \frac{1}{(2 \lambda)^{7/2}}  \,.
\eea

\section{Hermite-Gaussian integrals}
In the main text, we have defined the following Hermite-Gaussian integrals [\eq{ggmn}]: 
\be
\widetilde{gg}_{mn}(\lambda, \lambda^{'}, \tau_1, \tau_2)=\int d k_{1} d k_{2} k_{1}^{m}k_{2}^{n} 
{\rm e}^{-ik_{1} \tau_{1}} {\rm e}^{-ik_{2} \tau_{2}} {\rm e}^{-\gamma (k_{1}^{2}+ k_{2}^{2})} \,.
\ee
Here we evaluate them for $0 \leq m+n \leq 4$ with $m$ and $n$ being integer and $\gamma = \frac{1}{4 \lambda} +\frac{1}{4 \lambda^{'}} = \frac{\lambda + \lambda^{'}}{4 \lambda \lambda^{'}}$: 
\be
\widetilde{gg}_{00}(\lambda, \lambda', \tau_1, \tau_2) =  \frac{\pi}{\gamma} {\rm e}^{-\tau_{1}^{2}/4\gamma} {\rm e}^{-\tau_{2}^{2}/4\gamma} \,,
\label{tgg1}
\ee
\be
\widetilde{gg}_{10}(\lambda, \lambda', \tau_1, \tau_2) = \frac{\pi}{\gamma}\frac{-i\tau_{1}}{2\gamma} {\rm e}^{-\tau_{1}^{2}/4\gamma} {\rm e}^{-\tau_{2}^{2}/4\gamma} \,,
\ee
\be
\widetilde{gg}_{01}(\lambda, \lambda', \tau_1, \tau_2) = \frac{\pi}{\gamma}\frac{-i\tau_{2}}{2\gamma} {\rm e}^{-\tau_{1}^{2}/4\gamma} {\rm e}^{-\tau_{2}^{2}/4\gamma} \,,
\ee
\be
\widetilde{gg}_{20}(\lambda, \lambda', \tau_1, \tau_2) = \frac{\pi}{\gamma}\frac{1}{2\gamma} \left( 1- \frac{\tau_{1}^{2}}{2\gamma} \right){\rm  e}^{-\tau_{1}^{2}/4\gamma} {\rm e}^{-\tau_{2}^{2}/4\gamma} \,,
\ee
\be
\widetilde{gg}_{11}(\lambda, \lambda', \tau_1, \tau_2) = \frac{\pi}{\gamma} \frac{-\tau_{1}\tau_{2}}{4\gamma^{2}} {\rm  e}^{-\tau_{1}^{2}/4\gamma} {\rm e}^{-\tau_{2}^{2}/4\gamma} \,,
\ee
\be
\widetilde{gg}_{02}(\lambda, \lambda', \tau_1, \tau_2) = \frac{\pi}{\gamma}\frac{1}{2\gamma} \left( 1- \frac{\tau_{2}^{2}}{2\gamma} \right){\rm  e}^{-\tau_{1}^{2}/4\gamma} {\rm e}^{-\tau_{2}^{2}/4\gamma} \,,
\ee
\be
\widetilde{gg}_{30}(\lambda, \lambda', \tau_1, \tau_2) = \frac{\pi}{\gamma} \frac{-3 i\tau_{1}}{4\gamma^{2}} \left( 1- \frac{\tau_{1}^{2}}{6\gamma} \right) {\rm  e}^{-\tau_{1}^{2}/4\gamma} {\rm e}^{-\tau_{2}^{2}/4\gamma} \,,
\ee
\be
\widetilde{gg}_{21}(\lambda, \lambda', \tau_1, \tau_2) = \frac{\pi}{\gamma} \frac{-i\tau_{2}}{4\gamma^{2}} \left( 1- \frac{\tau_{1}^{2}}{2\gamma} \right)
{\rm  e}^{-\tau_{1}^{2}/4\gamma} {\rm e}^{-\tau_{2}^{2}/4\gamma} \,,
\ee
\be
\widetilde{gg}_{12}(\lambda, \lambda', \tau_1, \tau_2) =  \frac{\pi}{\gamma} \frac{-i\tau_{1}}{4\gamma^{2}} \left( 1- \frac{\tau_{2}^{2}}{2\gamma} \right)
{\rm  e}^{-\tau_{1}^{2}/4\gamma} {\rm e}^{-\tau_{2}^{2}/4\gamma} \,,
\ee
\be
\widetilde{gg}_{03}(\lambda, \lambda', \tau_1, \tau_2) = \frac{\pi}{\gamma} \frac{-3 i\tau_{2}}{4\gamma^{2}} \left( 1- \frac{\tau_{2}^{2}}{6\gamma} \right) {\rm  e}^{-\tau_{1}^{2}/4\gamma} {\rm e}^{-\tau_{2}^{2}/4\gamma} \,,
\ee
\be
\widetilde{gg}_{40}(\lambda, \lambda', \tau_1, \tau_2) =  \frac{\pi}{\gamma} \frac{3}{4\gamma^{2}} \left( 1- \frac{\tau_{1}^{2}}{\gamma}+\frac{\tau_{1}^{4}}{12\gamma^{2}} \right) {\rm  e}^{-\tau_{1}^{2}/4\gamma} {\rm e}^{-\tau_{2}^{2}/4\gamma} \,,
\ee
\be
\widetilde{gg}_{31}(\lambda, \lambda', \tau_1, \tau_2) = \frac{\pi}{\gamma} \frac{-3 \tau_{1} \tau_{2}}{8\gamma^{3}} \left( 1- \frac{\tau_{1}^{2}}{6\gamma} \right) {\rm  e}^{-\tau_{1}^{2}/4\gamma} {\rm e}^{-\tau_{2}^{2}/4\gamma} \,,
\ee
\be
\widetilde{gg}_{22}(\lambda, \lambda', \tau_1, \tau_2) = \frac{\pi}{\gamma} \frac{1}{4\gamma^{2}} \left( 1- \frac{\tau_{1}^{2}}{2\gamma} \right) \left( 1- \frac{\tau_{2}^{2}}{2\gamma}\right) {\rm  e}^{-\tau_{1}^{2}/4\gamma} {\rm e}^{-\tau_{2}^{2}/4\gamma} \,,
\ee
\be
\widetilde{gg}_{13}(\lambda, \lambda', \tau_1, \tau_2) = \frac{\pi}{\gamma} \frac{-3 \tau_{1} \tau_{2}}{8\gamma^{3}} \left( 1- \frac{\tau_{2}^{2}}{6\gamma} \right) {\rm  e}^{-\tau_{1}^{2}/4\gamma} {\rm e}^{-\tau_{2}^{2}/4\gamma} \,,
\ee
\be
\widetilde{gg}_{04}(\lambda, \lambda', \tau_1, \tau_2) =  \frac{\pi}{\gamma} \frac{3}{4\gamma^{2}} \left( 1- \frac{\tau_{2}^{2}}{\gamma}+\frac{\tau_{2}^{4}}{12\gamma^{2}} \right) {\rm  e}^{-\tau_{1}^{2}/4\gamma} {\rm e}^{-\tau_{2}^{2}/4\gamma} \,.
\label{tgg15}
\ee

\section{Coefficients of \boldmath{$A(\nu_{1},a_{1})$} and \boldmath{$B(k_{3}, \nu_{1},a_{1},\nu_{2},a_{2},\tau_{1}, \tau_{2}, \theta, \phi)$}}
To calculate the coefficients of \eq{I-C}, we use \eq{kxk1} and write $k_{x}=\sum_{j=1}^{3}u_{1j}k_{j}$, $k_{y}=\sum_{j=1}^{3}u_{2j}k_{j}$, $k_{z}=\sum_{j=1}^{3}u_{3j}k_{j}$. 

\subsection{For {$\nu_{1}=1s6, 2s3, 2s1$}} 
For $\nu_{1}=1s6, 2s3, 2s1$, 
\be
A(\nu_{1},a_{1})= w_{s}(\lambda_{\nu_{1}}^{a_{1}})\,.
\ee

When $\nu_{2}=1s6, 2s3, 2s1$, 
\be
B(k_{3}, \nu_{1},a_{1},\nu_{2},a_{2}, \tau_{1}, \tau_{2}, \theta, \phi)= w_{s}(\lambda_{\nu_{2}}^{a_{2}})gg_{00}(\lambda_{\nu_{1}}^{a1}, \lambda_{\nu_{2}}^{a2}, \tau_{1}, \tau_{2})\,.
\ee
To make expressions simpler, we will omit the arguments of $gg_{ij}$ on the right-hand side in the following---its arguments can be clear from the left-hand-side of expression. 
When $\nu_{2}=2p_{x}3, 2p_{x}1$, 
\be
B(k_{3}, \nu_{1},a_{1},\nu_{2},a_{2}, \tau_{1}, \tau_{2}, \theta, \phi)=-i w_{p}(\lambda_{\nu_{2}}^{a_{2}})(i u_{11}gg_{10} +i u_{12}gg_{01} + u_{13}k_{3}gg_{00})=-i w_{p}(\lambda_{\nu_{2}}^{a_{2}}) gg_{x} \,,
\ee
where we define the function $gg_{x}$, whose functional form is clear from the above equation. We also omit the arguments of $gg_{x}$ because it is the same as that of $gg_{ij}$; $\theta$ and $\phi$ dependence comes from $u_{ij}$ [see \eq{kxk1}].  Similarly when $\nu_{2}=2p_{y}3, 2p_{y}1$ and $\nu_{2}=2p_{z}3, 2p_{z}1$, we define the function $gg_{y}$ and $gg_{z}$, respectively, 
\be
B(k_{3}, \nu_{1},a_{1},\nu_{2},a_{2}, \tau_{1}, \tau_{2}, \theta, \phi)=-i w_{p}(\lambda_{\nu_{2}}^{a_{2}})(i u_{21}gg_{10} +i u_{22}gg_{01} + u_{23}k_{3}gg_{00})=-i w_{p}(\lambda_{\nu_{2}}^{a_{2}}) gg_{y} \,,
\ee
\be
B(k_{3}, \nu_{1},a_{1},\nu_{2},a_{2}, \tau_{1}, \tau_{2}, \theta, \phi)=-i w_{p}(\lambda_{\nu_{2}}^{a_{2}})(i u_{31}gg_{10} +i u_{32}gg_{01} + u_{33}k_{3}gg_{00})=-i w_{p}(\lambda_{\nu_{2}}^{a_{2}} )gg_{z} \,.
\ee
When $\nu_{2}=3d_{xx}$,  
\bea
&&B(k_{3}, \nu_{1},a_{1},\nu_{2},a_{2}, \tau_{1}, \tau_{2}, \theta, \phi)=w_{dd}(\lambda_{\nu_{2}}^{a_{2}}) [(2 \lambda_{\nu_{2}}^{a_{2}} - u_{13}^{2}k_{3}^{2})gg_{00} 
- u_{11}^{2}gg_{20} -2 u_{11}u_{12}gg_{11} \nonumber\\ 
&&\hspace{60mm}
-u_{12}^{2} gg_{02} -i2 u_{13}k_{3} ( u_{11} gg_{10} + u_{12} gg_{01})] \label{sdxx}\\
&&\hspace{45mm}=w_{dd}(\lambda_{\nu_{2}}^{a_{2}}) (2 \lambda_{\nu_{2}}^{a_{2}} gg_{00} - gg_{xx}) \,, 
\eea 
where we define the function $gg_{xx}$, whose functional form is clear from \eq{sdxx}. Similarly we obtain  $gg_{yy}$ and $gg_{zz}$, when $\nu_{2}=3d_{yy}$ and 3d$_{zz}$, respectively, by replacing $u_{1j} \rightarrow u_{2j}$ and 
  $u_{1j} \rightarrow u_{3j}$ in $gg_{xx}$ 
\be
B(k_{3}, \nu_{1},a_{1},\nu_{2},a_{2}, \tau_{1}, \tau_{2}, \theta, \phi)=w_{dd}(\lambda_{\nu_{2}}^{a_{2}}) (2 \lambda_{\nu_{2}}^{a_{2}} gg_{00} - gg_{yy}) \,,
\ee 
\be
B(k_{3}, \nu_{1},a_{1},\nu_{2},a_{2}, \tau_{1}, \tau_{2}, \theta, \phi)=w_{dd}(\lambda_{\nu_{2}}^{a_{2}}) (2 \lambda_{\nu_{2}}^{a_{2}} gg_{00} - gg_{zz}) \,.
\ee 
When $\nu_{2}=3d_{xy}$, 
\bea
&&B(k_{3}, \nu_{1},a_{1},\nu_{2},a_{2}, \tau_{1}, \tau_{2}, \theta, \phi)=-w_{d}(\lambda_{\nu_{2}}^{a_{2}}) [u_{11}u_{21}gg_{20} +u_{12}u_{22}gg_{02}+ u_{13}u_{23}k_{3}^{2}gg_{00} \nonumber  \\
&& \hspace{10mm}  +(u_{11}u_{22} + u_{12}u_{21})gg_{11}  + i (u_{11} u_{23}+ u_{13}u_{21})k_{3} gg_{10} + i (u_{12} u_{23}+ u_{13}u_{22})k_{3} gg_{01} \label{sdxy} \\
&&\hspace{45mm}=-w_{d}(\lambda_{\nu_{2}}^{a_{2}}) gg_{xy} \,, 
\eea 
where we have introduced a function $gg_{xy}$, whose functional form is clear from \eq{sdxy}.  

Similarly when $\nu_{2}=3d_{xz}$ and $3d_{yz}$, we replace $u_{2j}\rightarrow u_{3j}$ and $u_{1j} \rightarrow  u_{3j}$, respectively, 
\be
B(k_{3}, \nu_{1},a_{1},\nu_{2},a_{2}, \tau_{1}, \tau_{2}, \theta, \phi)=-w_{d}(\lambda_{\nu_{2}}^{a_{2}}) gg_{xz} \,, 
\ee 
\be
B(k_{3}, \nu_{1},a_{1},\nu_{2},a_{2}, \tau_{1}, \tau_{2}, \theta, \phi)=-w_{d}(\lambda_{\nu_{2}}^{a_{2}}) gg_{yz} \,.
\ee 

\subsection{For {$\nu_{1}=2p_{x}3$ and $2p_{x}1$}} 
For $\nu_{1}=2p_{x}3$ and $2p_{x}1$, 
\be
A(\nu_{1},a_{1})= iw_{p}(\lambda_{\nu_{1}}^{a_{1}}) \,.
\ee

When $\nu_{2}=1s6, 2s3, 2s1$, 
\be
B(k_{3}, \nu_{1},a_{1},\nu_{2},a_{2}, \tau_{1}, \tau_{2}, \theta, \phi) =w_{s}(\lambda_{\nu_{2}}^{a_{2}})gg_{x} \,.
\ee
When $\nu_{2}=2p_{x}3, 2p_{x}1$,
\be
B(k_{3}, \nu_{1},a_{1},\nu_{2},a_{2}, \tau_{1}, \tau_{2}, \theta, \phi)=-i w_{p}(\lambda_{\nu_{2}}^{a_{2}}) gg_{xx}\,.
\ee 
When $\nu_{2}=2p_{y}3, 2p_{y}1$,
\be
B(k_{3}, \nu_{1},a_{1},\nu_{2},a_{2}, \tau_{1}, \tau_{2}, \theta, \phi)=-i w_{p}(\lambda_{\nu_{2}}^{a_{2}}) gg_{xy}\,.
\ee 
When $\nu_{2}=2p_{z}3, 2p_{z}1$,
\be
B(k_{3}, \nu_{1},a_{1},\nu_{2},a_{2}, \tau_{1}, \tau_{2}, \theta, \phi)=-i w_{p}(\lambda_{\nu_{2}}^{a_{2}}) gg_{xz}\,.
\ee 
When $\nu_{2}=3d_{xx}$,  
\be
B(k_{3}, \nu_{1},a_{1},\nu_{2},a_{2}, \tau_{1}, \tau_{2}, \theta, \phi)=w_{dd}(\lambda_{\nu_{2}}^{a_{2}}) (2 \lambda_{\nu_{2}}^{a_{2}} gg_{x} - gg_{xxx}) \, .
\ee
To define $gg_{xxx}$, we first define a general formula of $gg_{xyz}$ as
\bea
&&gg_{xyz}= u_{13}u_{23}u_{33}k_{3}^{3} gg_{00} + (u_{13}u_{21} u_{31}+ u_{11}u_{23} u_{31} +u_{11}u_{21} u_{33})k_{3}gg_{20} \nonumber \\
&& \hspace{10mm} +(u_{13}u_{22}u_{31}+ u_{12}u_{23}u_{31}+u_{13}u_{21}u_{32} +
u_{11}u_{23}u_{32}+ u_{12}u_{21}u_{33}+u_{11}u_{22}u_{33})k_{3} gg_{11}  \nonumber \\ 
&&\hspace{10mm} +(u_{13}u_{22}u_{32} + u_{12}u_{23}u_{32} +u_{12}u_{22}u_{33}) k_{3}gg_{02} \nonumber \\
&&\hspace{10mm}+i (u_{13}u_{23}u_{31} + u_{13} u_{21}u_{33} +u_{11} u_{23}u_{33})k_{3}^{2} gg_{10} \nonumber \\
&& \hspace{10mm} + i (u_{13}u_{23}u_{32} + u_{13} u_{22}u_{33} +u_{12} u_{23}u_{33})k_{3}^{2} gg_{01} \nonumber \\
&&\hspace{10mm}+ iu_{11}u_{21}u_{31} gg_{30} +i (u_{12}u_{21}u_{31} + u_{11}u_{22}u_{31} + u_{11}u_{21}u_{32}) gg_{21} \nonumber \\
&&\hspace{10mm} + i (u_{12}u_{22}u_{31} + u_{12}u_{21}u_{32} + u_{11}u_{22}u_{32}) gg_{12} 
+ i u_{12}u_{22} u_{32}gg_{03} \,.
\label{ggxyz}
\eea
$gg_{xxx}$ is obtained by replacing $(u_{2j}, u_{3j}) \rightarrow (u_{1j}, u_{1j})$.

When $\nu_{2}=3d_{yy}$,
\be
B(k_{3}, \nu_{1},a_{1},\nu_{2},a_{2}, \tau_{1}, \tau_{2}, \theta, \phi)=w_{dd}(\lambda_{\nu_{2}}^{a_{2}}) (2 \lambda_{\nu_{2}}^{a_{2}} gg_{x} - gg_{xyy}) \, ,
\ee
where $gg_{xyy}$ is obtained by  replacing $u_{3j} \rightarrow u_{2j}$ in \eq{ggxyz}.

When $\nu_{2}=3d_{zz}$,
\be
B(k_{3}, \nu_{1},a_{1},\nu_{2},a_{2}, \tau_{1}, \tau_{2}, \theta, \phi)=w_{dd}(\lambda_{\nu_{2}}^{a_{2}}) (2 \lambda_{\nu_{2}}^{a_{2}} gg_{x} - gg_{xzz}) \, ,
\ee
where $gg_{xzz}$ is obtained by  replacing $u_{2j} \rightarrow u_{3j}$ in \eq{ggxyz}.

When $\nu_{2}=3d_{xy}$, 
\be
B(k_{3}, \nu_{1},a_{1},\nu_{2},a_{2}, \tau_{1}, \tau_{2}, \theta, \phi)=-w_{d}(\lambda_{\nu_{2}}^{a_{2}}) gg_{xxy}\, ,
\ee
where $gg_{xxy}$ is obtained by  replacing $u_{3j} \rightarrow u_{1j}$ in \eq{ggxyz}.

When $\nu_{2}=3d_{xz}$, 
\be
B(k_{3}, \nu_{1},a_{1},\nu_{2},a_{2}, \tau_{1}, \tau_{2}, \theta, \phi)=-w_{d}(\lambda_{\nu_{2}}^{a_{2}}) gg_{xxz}\, ,
\ee
where $g_{xxz}$ is obtained by  replacing $u_{2j} \rightarrow u_{1j}$ in \eq{ggxyz}. 

When $\nu_{2}=3d_{yz}$, 
\be
B(k_{3}, \nu_{1},a_{1},\nu_{2},a_{2}, \tau_{1}, \tau_{2}, \theta, \phi)=-w_{d}(\lambda_{\nu_{2}}^{a_{2}}) gg_{xyz}\, .
\ee

\subsection{For {$\nu_{1}=2p_{y}3$ and $2p_{y}1$}} 
For $\nu_{1}=2p_{y}3$ and $2p_{y}1$, 
\be
A(\nu_{1},a_{1})= iw_{p}(\lambda_{\nu_{1}}^{a_{1}}) \,.
\ee

When $\nu_{2}=1s6, 2s3, 2s1$, 
\be
B(k_{3}, \nu_{1},a_{1},\nu_{2},a_{2}, \tau_{1}, \tau_{2}, \theta, \phi)= w_{s}(\lambda_{\nu_{2}}^{a_{2}})gg_{y} \,.
\ee
When $\nu_{2}=2p_{x}3, 2p_{x}1$,
\be
B(k_{3}, \nu_{1},a_{1},\nu_{2},a_{2}, \tau_{1}, \tau_{2}, \theta, \phi)=-i w_{p}(\lambda_{\nu_{2}}^{a_{2}}) gg_{xy}\,.
\ee 
When $\nu_{2}=2p_{y}3, 2p_{y}1$,
\be
B(k_{3}, \nu_{1},a_{1},\nu_{2},a_{2}, \tau_{1}, \tau_{2}, \theta, \phi)=-i w_{p}(\lambda_{\nu_{2}}^{a_{2}}) gg_{yy}\,.
\ee 
When $\nu_{2}=2p_{z}3, 2p_{z}1$,
\be
B(k_{3}, \nu_{1},a_{1},\nu_{2},a_{2}, \tau_{1}, \tau_{2}, \theta, \phi)=-i w_{p}(\lambda_{\nu_{2}}^{a_{2}}) gg_{yz}\,.
\ee 
When $\nu_{2}=3d_{xx}$,  
\be
B(k_{3}, \nu_{1},a_{1},\nu_{2},a_{2}, \tau_{1}, \tau_{2}, \theta, \phi)=w_{dd}(\lambda_{\nu_{2}}^{a_{2}}) (2 \lambda_{\nu_{2}}^{a_{2}} gg_{y} - gg_{xxy})\, .
\ee
When $\nu_{2}=3d_{yy}$,
\be
B(k_{3}, \nu_{1},a_{1},\nu_{2},a_{2}, \tau_{1}, \tau_{2}, \theta, \phi)=w_{dd}(\lambda_{\nu_{2}}^{a_{2}}) (2 \lambda_{\nu_{2}}^{a_{2}} gg_{y} - gg_{yyy})\, ,
\ee
where $gg_{yyy}$ is obtained by replacing $(u_{1j}, u_{3j}) \rightarrow (u_{2j}, u_{2j})$ in \eq{ggxyz}.  

When $\nu_{2}=3d_{zz}$,
\be
B(k_{3}, \nu_{1},a_{1},\nu_{2},a_{2}, \tau_{1}, \tau_{2}, \theta, \phi)=w_{dd}(\lambda_{\nu_{2}}^{a_{2}}) (2 \lambda_{\nu_{2}}^{a_{2}} gg_{y} - gg_{yzz})\, ,
\ee
where $gg_{yzz}$ is obtained by replacing $u_{1j} \rightarrow  u_{3j}$ in \eq{ggxyz}.

When $\nu_{2}=3d_{xy}$, 
\be
B(k_{3}, \nu_{1},a_{1},\nu_{2},a_{2}, \tau_{1}, \tau_{2}, \theta, \phi)=-w_{d}(\lambda_{\nu_{2}}^{a_{2}}) gg_{xyy}\, .
\ee
When $\nu_{2}=3d_{xz}$, 
\be
B(k_{3}, \nu_{1},a_{1},\nu_{2},a_{2}, \tau_{1}, \tau_{2}, \theta, \phi)=-w_{d}(\lambda_{\nu_{2}}^{a_{2}}) gg_{xyz}\, .
\ee
When $\nu_{2}=3d_{yz}$, 
\be
B(k_{3}, \nu_{1},a_{1},\nu_{2},a_{2}, \tau_{1}, \tau_{2}, \theta, \phi)=-w_{d}(\lambda_{\nu_{2}}^{a_{2}}) gg_{yyz}\, ,
\ee
where $gg_{yyz}$ is obtained by replacing $u_{1j} \rightarrow  u_{2j}$ in \eq{ggxyz}.

\subsection{For {$\nu_{1}=2p_{z}3$ and $2p_{z}1$}} 
For $\nu_{1}=2p_{z}3$ and $2p_{z}1$, 
\be
A(\nu_{1},a_{1})= iw_{p}(\lambda_{\nu_{1}}^{a_{1}}) \,.
\ee

When $\nu_{2}=1s6, 2s3, 2s1$, 
\be
B(k_{3}, \nu_{1},a_{1},\nu_{2},a_{2}, \tau_{1}, \tau_{2}, \theta, \phi)= w_{s}(\lambda_{\nu_{2}}^{a_{2}})gg_{z}\,.
\ee
When $\nu_{2}=2p_{x}3, 2p_{x}1$,
\be
B(k_{3}, \nu_{1},a_{1},\nu_{2},a_{2}, \tau_{1}, \tau_{2}, \theta, \phi)=-i w_{p}(\lambda_{\nu_{2}}^{a_{2}}) gg_{xz}\,.
\ee 
When $\nu_{2}=2p_{y}3, 2p_{y}1$,
\be
B(k_{3}, \nu_{1},a_{1},\nu_{2},a_{2}, \tau_{1}, \tau_{2}, \theta, \phi)=-i w_{p}(\lambda_{\nu_{2}}^{a_{2}}) gg_{yz}\,.
\ee 
When $\nu_{2}=2p_{z}3, 2p_{z}1$,
\be
B(k_{3}, \nu_{1},a_{1},\nu_{2},a_{2}, \tau_{1}, \tau_{2}, \theta, \phi)=-i w_{p}(\lambda_{\nu_{2}}^{a_{2}}) gg_{zz}\,.
\ee 
When $\nu_{2}=3d_{xx}$,  
\be
B(k_{3}, \nu_{1},a_{1},\nu_{2},a_{2}, \tau_{1}, \tau_{2}, \theta, \phi)=w_{dd}(\lambda_{\nu_{2}}^{a_{2}}) (2 \lambda_{\nu_{2}}^{a_{2}} gg_{z} - gg_{xxz})\, .
\ee
When $\nu_{2}=3d_{yy}$,
\be
B(k_{3}, \nu_{1},a_{1},\nu_{2},a_{2}, \tau_{1}, \tau_{2}, \theta, \phi)=w_{dd}(\lambda_{\nu_{2}}^{a_{2}}) (2 \lambda_{\nu_{2}}^{a_{2}} gg_{z} - gg_{yyz})\, . 
\ee
When $\nu_{2}=3d_{zz}$,
\be
B(k_{3}, \nu_{1},a_{1},\nu_{2},a_{2}, \tau_{1}, \tau_{2}, \theta, \phi)=w_{dd}(\lambda_{\nu_{2}}^{a_{2}}) (2 \lambda_{\nu_{2}}^{a_{2}} gg_{z} - gg_{zzz})\, .
\ee
Here $gg_{zzz}$ is obtained by replacing $(u_{1j},u_{2j}) \rightarrow (u_{3j},u_{3j})$ in \eq{ggxyz}.

When $\nu_{2}=3d_{xy}$, 
\be
B(k_{3}, \nu_{1},a_{1},\nu_{2},a_{2}, \tau_{1}, \tau_{2}, \theta, \phi)=-w_{d}(\lambda_{\nu_{2}}^{a_{2}}) gg_{xyz}\, .
\ee
When $\nu_{2}=3d_{xz}$, 
\be
B(k_{3}, \nu_{1},a_{1},\nu_{2},a_{2}, \tau_{1}, \tau_{2}, \theta, \phi)=-w_{d}(\lambda_{\nu_{2}}^{a_{2}}) gg_{xzz}\, .
\ee
When $\nu_{2}=3d_{yz}$, 
\be
B(k_{3}, \nu_{1},a_{1},\nu_{2},a_{2}, \tau_{1}, \tau_{2}, \theta, \phi)=-w_{d}(\lambda_{\nu_{2}}^{a_{2}}) gg_{yzz}\, .
\ee

\subsection{For {$\nu_{1}=3d_{xx}$}} 
For $\nu_{1}=3d_{xx}$, 
\be
A(\nu_{1},a_{1})= w_{dd}(\lambda_{\nu_{1}}^{a_{1}}) \,.
\ee

When $\nu_{2}=1s6, 2s3, 2s1$, 
\be
B(k_{3}, \nu_{1},a_{1},\nu_{2},a_{2}, \tau_{1}, \tau_{2}, \theta, \phi)= w_{s}(\lambda_{\nu_{2}}^{a_{2}})( 2\lambda_{\nu_{1}}^{a_{1}}gg_{00} - gg_{xx}) \,.
\ee
When $\nu_{2}=2p_{x}3, 2p_{x}1$,
\be
B(k_{3}, \nu_{1},a_{1},\nu_{2},a_{2}, \tau_{1}, \tau_{2}, \theta, \phi)=-i w_{p}(\lambda_{\nu_{2}}^{a_{2}}) (2\lambda_{\nu_{1}}^{a_{1}}gg_{x} - gg_{xxx})  \,.
\ee 
When $\nu_{2}=2p_{y}3, 2p_{y}1$,
\be
B(k_{3}, \nu_{1},a_{1},\nu_{2},a_{2}, \tau_{1}, \tau_{2}, \theta, \phi)=-i w_{p}(\lambda_{\nu_{2}}^{a_{2}}) (2\lambda_{\nu_{1}}^{a_{1}}gg_{y} - gg_{xxy}) \,.
\ee 
When $\nu_{2}=2p_{z}3, 2p_{z}1$,
\be
B(k_{3}, \nu_{1},a_{1},\nu_{2},a_{2}, \tau_{1}, \tau_{2}, \theta, \phi)=-i w_{p}(\lambda_{\nu_{2}}^{a_{2}}) (2\lambda_{\nu_{1}}^{a_{1}}gg_{z} - gg_{xxz}) \,.
\ee 
When $\nu_{2}=3d_{xx}$,  
\be
B(k_{3}, \nu_{1},a_{1},\nu_{2},a_{2}, \tau_{1}, \tau_{2}, \theta, \phi)=w_{dd}(\lambda_{\nu_{2}}^{a_{2}}) [4 \lambda_{\nu_{1}}^{a_{1}}\lambda_{\nu_{2}}^{a_{2}} gg_{00}
-2(\lambda_{\nu_{1}}^{a_{1}}+ \lambda_{\nu_{2}}^{a_{2}})gg_{xx}  + gg_{xxxx}]\, .
\ee
To define $gg_{xxxx}$, we first define a general expression of $gg_{xxyz}$: 
\bea
&&gg_{xxyz}=u_{11}^{2}u_{21}u_{31}gg_{40}+u_{12}^{2}u_{22}u_{32}gg_{04}+u_{13}^{2}u_{23}u_{33}k_{3}^{4}gg_{00}  \nonumber\\
&& \hspace{15mm} + (2u_{11}u_{12}u_{21}u_{31}+u_{11}^{2}u_{22}u_{31}+u_{11}^{2}u_{21}u_{32})gg_{31} \nonumber \\
&&\hspace{15mm}  + (u_{12}^{2}u_{21}u_{31}+2u_{11}u_{12}u_{22}u_{31}+2u_{11}u_{12}u_{21}u_{32}+u_{11}^{2}u_{22}u_{32})gg_{22}\nonumber \\
&&\hspace{15mm} +(u_{12}^{2}u_{22}u_{31}+u_{12}^{2}u_{21}u_{32}+2u_{11}u_{12}u_{22}u_{32})gg_{13}\nonumber \\
&&\hspace{15mm}  +(u_{13}^{2}u_{22}u_{31}+2u_{12}u_{13}u_{23}u_{31}+u_{13}^{2}u_{21}u_{32}+2u_{11} u_{13}u_{23}u_{32}+2u_{12}u_{13}u_{21}u_{33}
\nonumber \\ 
&&\hspace{30mm} + 2u_{11}u_{13}u_{22}u_{33}+2u_{11}u_{12}u_{23}u_{33})k_{3}^{2}gg_{11}\nonumber \\
&& \hspace{15mm} +(u_{13}^{2}u_{22}u_{32} + 2u_{12}u_{13}u_{23}u_{32}+2u_{12}u_{13}u_{22}u_{33}+u_{12}^{2}u_{23}u_{33})k_{3}^{2}gg_{02}\nonumber \\
&&\hspace{15mm} +(u_{13}^{2}u_{21}u_{31}+2u_{11}u_{13}u_{23}u_{31}+2u_{11}u_{13}u_{21}u_{33}+u_{11}^{2}u_{23}u_{33})k_{3}^{2}gg_{20}\nonumber \\
&& \hspace{15mm} +i(2u_{11}u_{13}u_{21}u_{31}+u_{11}^{2}u_{23}u_{31}+u_{11}^{2}u_{21}u_{33})k_{3}gg_{30}\nonumber \\
&&\hspace{15mm}  +i(2u_{12}u_{13}u_{21}u_{31} + 2u_{11}u_{13}u_{22}u_{31}+2u_{11}u_{12}u_{23}u_{31}+2u_{11}u_{13}u_{21}u_{32} \nonumber \\
&& \hspace{30mm}+u_{11}^{2}u_{23}u_{32}+ 2u_{11}u_{12}u_{21}u_{33} +u_{11}^{2}u_{22}u_{33})k_{3}gg_{21}\nonumber \\
&&\hspace{15mm} +i(2u_{12}u_{13}u_{22}u_{31}+u_{12}^{2}u_{23}u_{31}+2u_{12}u_{13}u_{21}u_{32}+2u_{11}u_{13}u_{22}u_{32} \nonumber  \\
&&\hspace{30mm} +2u_{11}u_{12}u_{23}u_{32}+u_{12}^{2}u_{21}u_{33}+2u_{11}u_{12}u_{22}u_{33})k_{3}gg_{12}\nonumber \\
&&\hspace{15mm} + i(2u_{12}u_{13}u_{22}u_{32}+u_{12}^{2}u_{23}u_{32}+u_{12}^{2}u_{22}u_{33})k_{3}gg_{03} \nonumber \\
&&\hspace{15mm} +i (u_{13}^{2}u_{23}u_{32}+u_{13}^{2}u_{22}u_{33}+2u_{12}u_{13}u_{23}u_{33})k_{3}^{3}gg_{01}
\nonumber \\
&&\hspace{15mm} +i (u_{13}^{2}u_{23}u_{31}+u_{13}^{2}u_{21}u_{33}+ 2 u_{11}u_{13}u_{23}u_{33}) k_{3}^{3}gg_{10}\,.
\label{ggxxyz}
\eea
$gg_{xxxx}$ is obtained by replacing $(u_{2j}, u_{3j}) \rightarrow (u_{1j}, u_{1j})$.

When $\nu_{2}=3d_{yy}$,  
\be
B(k_{3}, \nu_{1},a_{1},\nu_{2},a_{2}, \tau_{1}, \tau_{2}, \theta, \phi)=w_{dd}(\lambda_{\nu_{2}}^{a_{2}}) (4 \lambda_{\nu_{1}}^{a_{1}}\lambda_{\nu_{2}}^{a_{2}} gg_{00}
-2\lambda_{\nu_{2}}^{a_{2}}gg_{xx} -2\lambda_{\nu_{1}}^{a_{1}}gg_{yy}  + gg_{xxyy}) \, ,
\ee
where $gg_{xxyy}$ is obtained by $u_{3j} \rightarrow u_{2j}$ in \eq{ggxxyz}.

When $\nu_{2}=3d_{zz}$,  
\be
B(k_{3}, \nu_{1},a_{1},\nu_{2},a_{2}, \tau_{1}, \tau_{2}, \theta, \phi)=w_{dd}(\lambda_{\nu_{2}}^{a_{2}}) (4 \lambda_{\nu_{1}}^{a_{1}}\lambda_{\nu_{2}}^{a_{2}} gg_{00}
-2\lambda_{\nu_{2}}^{a_{2}}gg_{xx} -2\lambda_{\nu_{1}}^{a_{1}}gg_{zz}  + gg_{xxzz}) \, ,
\ee
where $gg_{xxzz}$ is obtained by replacing $u_{2j} \rightarrow u_{3j}$ in \eq{ggxxyz}. 

When $\nu_{2}=3d_{xy}$, 
\be
B(k_{3}, \nu_{1},a_{1},\nu_{2},a_{2}, \tau_{1}, \tau_{2}, \theta, \phi)=-w_{d}(\lambda_{\nu_{2}}^{a_{2}})(2 \lambda_{\nu_{1}}^{a_{1}} gg_{xy} -gg_{xxxy} )\,,
\ee 
where $gg_{xxxy}$ is obtained by replacing $u_{3j} \rightarrow u_{1j}$ in \eq{ggxxyz}.

When $\nu_{2}=3d_{xz}$, 
\be
B(k_{3}, \nu_{1},a_{1},\nu_{2},a_{2}, \tau_{1}, \tau_{2}, \theta, \phi)=-w_{d}(\lambda_{\nu_{2}}^{a_{2}}) ( 2 \lambda_{\nu_{1}}^{a_{1}} gg_{xz}- gg_{xxxz}) \,,
\ee
where $gg_{xxxz}$ is obtained by replacing $u_{2j}\rightarrow u_{1j}$ in \eq{ggxxyz}. 

When $\nu_{2}=3d_{yz}$, 
\be
B(k_{3}, \nu_{1},a_{1},\nu_{2},a_{2}, \tau_{1}, \tau_{2}, \theta, \phi)=-w_{d}(\lambda_{\nu_{2}}^{a_{2}})(2 \lambda_{\nu_{1}}^{a_{1}} gg_{yz}- gg_{xxyz}) \,.
\ee

\subsection{For {$\nu_{1}=3d_{yy}$}} 
For $\nu_{1}=3d_{yy}$, 
\be
A(\nu_{1},a_{1})= w_{dd}(\lambda_{\nu_{1}}^{a_{1}}) \,.
\ee

When $\nu_{2}=1s6, 2s3, 2s1$, 
\be
B(k_{3}, \nu_{1},a_{1},\nu_{2},a_{2}, \tau_{1}, \tau_{2}, \theta, \phi)= w_{s}(\lambda_{\nu_{2}}^{a_{2}})( 2\lambda_{\nu_{1}}^{a_{1}}gg_{00} - gg_{yy}) \,.
\ee
When $\nu_{2}=2p_{x}3, 2p_{x}1$,
\be
B(k_{3}, \nu_{1},a_{1},\nu_{2},a_{2}, \tau_{1}, \tau_{2}, \theta, \phi)=-i w_{p}(\lambda_{\nu_{2}}^{a_{2}}) (2\lambda_{\nu_{1}}^{a_{1}}gg_{x} - gg_{xyy})  \,.
\ee 
When $\nu_{2}=2p_{y}3, 2p_{y}1$,
\be
B(k_{3}, \nu_{1},a_{1},\nu_{2},a_{2}, \tau_{1}, \tau_{2}, \theta, \phi)=-i w_{p}(\lambda_{\nu_{2}}^{a_{2}}) (2\lambda_{\nu_{1}}^{a_{1}}gg_{y} - gg_{yyy}) \,.
\ee 
When $\nu_{2}=2p_{z}3, 2p_{z}1$,
\be
B(k_{3}, \nu_{1},a_{1},\nu_{2},a_{2}, \tau_{1}, \tau_{2}, \theta, \phi)=-i w_{p}(\lambda_{\nu_{2}}^{a_{2}}) (2\lambda_{\nu_{1}}^{a_{1}}gg_{z} - gg_{yyz}) \,.
\ee 
When $\nu_{2}=3d_{xx}$,  
\be
B(k_{3}, \nu_{1},a_{1},\nu_{2},a_{2}, \tau_{1}, \tau_{2}, \theta, \phi)=w_{dd}(\lambda_{\nu_{2}}^{a_{2}}) (4 \lambda_{\nu_{1}}^{a_{1}}\lambda_{\nu_{2}}^{a_{2}} gg_{00}
-2\lambda_{\nu_{1}}^{a_{1}}gg_{xx} -2 \lambda_{\nu_{2}}^{a_{2}}gg_{yy}  + gg_{xxyy})\, .
\ee
When $\nu_{2}=3d_{yy}$,  
\be
B(k_{3}, \nu_{1},a_{1},\nu_{2},a_{2}, \tau_{1}, \tau_{2}, \theta, \phi)=w_{dd}(\lambda_{\nu_{2}}^{a_{2}}) [4 \lambda_{\nu_{1}}^{a_{1}}\lambda_{\nu_{2}}^{a_{2}} gg_{00}
-2(\lambda_{\nu_{1}}^{a_{1}} + \lambda_{\nu_{2}}^{a_{2}})gg_{yy}  + gg_{yyyy}] \, , 
\ee
where $gg_{yyyy}$ is obtained by replacing $(u_{1j},u_{3j}) \rightarrow (u_{2j}, u_{2j})$ in \eq{ggxxyz}.

When $\nu_{2}=3d_{zz}$,  
\be
B(k_{3}, \nu_{1},a_{1},\nu_{2},a_{2}, \tau_{1}, \tau_{2}, \theta, \phi)=w_{dd}(\lambda_{\nu_{2}}^{a_{2}}) (4 \lambda_{\nu_{1}}^{a_{1}}\lambda_{\nu_{2}}^{a_{2}} gg_{00}
-2\lambda_{\nu_{2}}^{a_{2}}gg_{yy} -2\lambda_{\nu_{1}}^{a_{1}}gg_{zz}  + gg_{yyzz}) \, ,
\ee
where $gg_{yyzz}$ is obtained by replacing $(u_{1j},u_{2j}) \rightarrow (u_{2j}, u_{3j})$ in \eq{ggxxyz}.

When $\nu_{2}=3d_{xy}$, 
\be
B(k_{3}, \nu_{1},a_{1},\nu_{2},a_{2}, \tau_{1}, \tau_{2}, \theta, \phi)=-w_{d}(\lambda_{\nu_{2}}^{a_{2}})(2 \lambda_{\nu_{1}}^{a_{1}} gg_{xy} -gg_{xyyy} )\,,
\ee 
where $gg_{xyyy}$ is obtained by replacing $(u_{1j},u_{3j}) \rightarrow (u_{2j}, u_{1j})$ in \eq{ggxxyz}.

When $\nu_{2}=3d_{xz}$, 
\be
B(k_{3}, \nu_{1},a_{1},\nu_{2},a_{2}, \tau_{1}, \tau_{2}, \theta, \phi)=-w_{d}(\lambda_{\nu_{2}}^{a_{2}})( 2 \lambda_{\nu_{1}}^{a_{1}} gg_{xz}- gg_{xyyz}) \,,
\ee
where $gg_{xyyz}$ is obtained by replacing $(u_{1j},u_{2j}) \rightarrow (u_{2j}, u_{1j})$ in \eq{ggxxyz}.

When $\nu_{2}=3d_{yz}$, 
\be
B(k_{3}, \nu_{1},a_{1},\nu_{2},a_{2}, \tau_{1}, \tau_{2}, \theta, \phi)=-w_{d}(\lambda_{\nu_{2}}^{a_{2}})( 2\lambda_{\nu_{1}}^{a_{1}} gg_{yz}- gg_{yyyz}) \,,
\ee
where $gg_{yyyz}$ is obtained by replacing $u_{1j} \rightarrow u_{2j}$ in \eq{ggxxyz}.

\subsection{For {$\nu_{1}=3d_{zz}$}} 
For $\nu_{1}=3d_{zz}$, 
\be
A(\nu_{1},a_{1})= w_{dd}(\lambda_{\nu_{1}}^{a_{1}}) \,.
\ee

When $\nu_{2}=1s6, 2s3, 2s1$, 
\be
B(k_{3}, \nu_{1},a_{1},\nu_{2},a_{2}, \tau_{1}, \tau_{2}, \theta, \phi)= w_{s}(\lambda_{\nu_{2}}^{a_{2}})( 2\lambda_{\nu_{1}}^{a_{1}}gg_{00} - gg_{zz}) \,.
\ee
When $\nu_{2}=2p_{x}3, 2p_{x}1$,
\be
B(k_{3}, \nu_{1},a_{1},\nu_{2},a_{2}, \tau_{1}, \tau_{2}, \theta, \phi)=-i w_{p}(\lambda_{\nu_{2}}^{a_{2}}) (2\lambda_{\nu_{1}}^{a_{1}}gg_{x} - gg_{xzz})  \,.
\ee 
When $\nu_{2}=2p_{y}3, 2p_{y}1$,
\be
B(k_{3}, \nu_{1},a_{1},\nu_{2},a_{2}, \tau_{1}, \tau_{2}, \theta, \phi)=-i w_{p}(\lambda_{\nu_{2}}^{a_{2}}) (2\lambda_{\nu_{1}}^{a_{1}}gg_{y} - gg_{yzz}) \,.
\ee 
When $\nu_{2}=2p_{z}3, 2p_{z}1$,
\be
B(k_{3}, \nu_{1},a_{1},\nu_{2},a_{2}, \tau_{1}, \tau_{2}, \theta, \phi)=-i w_{p}(\lambda_{\nu_{2}}^{a_{2}}) (2\lambda_{\nu_{1}}^{a_{1}}gg_{z} - gg_{zzz}) \,.
\ee 
When $\nu_{2}=3d_{xx}$,  
\be
B(k_{3}, \nu_{1},a_{1},\nu_{2},a_{2}, \tau_{1}, \tau_{2}, \theta, \phi)=w_{dd}(\lambda_{\nu_{2}}^{a_{2}}) (4 \lambda_{\nu_{1}}^{a_{1}}\lambda_{\nu_{2}}^{a_{2}} gg_{00}
-2\lambda_{\nu_{1}}^{a_{1}}gg_{xx} -2 \lambda_{\nu_{2}}^{a_{2}}gg_{zz}  + gg_{xxzz})\, .
\ee
When $\nu_{2}=3d_{yy}$,  
\be
B(k_{3}, \nu_{1},a_{1},\nu_{2},a_{2}, \tau_{1}, \tau_{2}, \theta, \phi)=w_{dd}(\lambda_{\nu_{2}}^{a_{2}}) (4 \lambda_{\nu_{1}}^{a_{1}}\lambda_{\nu_{2}}^{a_{2}} gg_{00}
-2\lambda_{\nu_{1}}^{a_{1}}gg_{yy} -2 \lambda_{\nu_{2}}^{a_{2}}gg_{zz} + gg_{yyzz}) \, .
\ee
When $\nu_{2}=3d_{zz}$,  
\be
B(k_{3}, \nu_{1},a_{1},\nu_{2},a_{2}, \tau_{1}, \tau_{2}, \theta, \phi)=w_{dd}(\lambda_{\nu_{2}}^{a_{2}}) [4 \lambda_{\nu_{1}}^{a_{1}}\lambda_{\nu_{2}}^{a_{2}} gg_{00}
-2(\lambda_{\nu_{1}}^{a_{1}} + \lambda_{\nu_{2}}^{a_{2}})gg_{zz}  + gg_{zzzz}] \, ,
\ee
where $gg_{zzzz}$ is obtained by replacing $(u_{1j},u_{2j}) \rightarrow (u_{3j}, u_{3j})$ in \eq{ggxxyz}.

When $\nu_{2}=3d_{xy}$, 
\be
B(k_{3}, \nu_{1},a_{1},\nu_{2},a_{2}, \tau_{1}, \tau_{2}, \theta, \phi)=-w_{d}(\lambda_{\nu_{2}}^{a_{2}})(2 \lambda_{\nu_{1}}^{a_{1}} gg_{xy} -gg_{xyzz} )\,,
\ee 
where $gg_{xyzz}$ is obtained by replacing $(u_{1j},u_{3j}) \rightarrow (u_{3j}, u_{1j})$ in \eq{ggxxyz}.

When $\nu_{2}=3d_{xz}$, 
\be
B(k_{3}, \nu_{1},a_{1},\nu_{2},a_{2}, \tau_{1}, \tau_{2}, \theta, \phi)=-w_{d}(\lambda_{\nu_{2}}^{a_{2}})( 2 \lambda_{\nu_{1}}^{a_{1}} gg_{xz}- gg_{xzzz}) \,,
\ee
where $gg_{xzzz}$ is obtained by replacing $(u_{1j},u_{2j}) \rightarrow (u_{3j}, u_{1j})$ in \eq{ggxxyz}.

When $\nu_{2}=3d_{yz}$, 
\be
B(k_{3}, \nu_{1},a_{1},\nu_{2},a_{2}, \tau_{1}, \tau_{2}, \theta, \phi)=-w_{d}(\lambda_{\nu_{2}}^{a_{2}})( 2\lambda_{\nu_{1}}^{a_{1}} gg_{yz}- gg_{yzzz}) \,,
\ee
where $gg_{yzzz}$ is obtained by replacing $u_{1j} \rightarrow u_{3j}$ in \eq{ggxxyz}.

\subsection{For {$\nu_{1}=3d_{xy}$}} 
For $\nu_{1}=3d_{xy}$, 
\be
A(\nu_{1},a_{1})= -w_{d}(\lambda_{\nu_{1}}^{a_{1}}) \,.
\ee

When $\nu_{2}=1s6, 2s3, 2s1$, 
\be
B(k_{3}, \nu_{1},a_{1},\nu_{2},a_{2}, \tau_{1}, \tau_{2}, \theta, \phi)= w_{s}(\lambda_{\nu_{2}}^{a_{2}}) gg_{xy} \,.
\ee
When $\nu_{2}=2p_{x}3, 2p_{x}1$,
\be
B(k_{3}, \nu_{1},a_{1},\nu_{2},a_{2}, \tau_{1}, \tau_{2}, \theta, \phi)=-i w_{p}(\lambda_{\nu_{2}}^{a_{2}}) gg_{xxy}  \,.
\ee 
When $\nu_{2}=2p_{y}3, 2p_{y}1$,
\be
B(k_{3}, \nu_{1},a_{1},\nu_{2},a_{2}, \tau_{1}, \tau_{2}, \theta, \phi)=-i w_{p}(\lambda_{\nu_{2}}^{a_{2}}) gg_{xyy} \,.
\ee 
When $\nu_{2}=2p_{z}3, 2p_{z}1$,
\be
B(k_{3}, \nu_{1},a_{1},\nu_{2},a_{2}, \tau_{1}, \tau_{2}, \theta, \phi)=-i w_{p}(\lambda_{\nu_{2}}^{a_{2}}) gg_{xyz} \,.
\ee 
When $\nu_{2}=3d_{xx}$,  
\be
B(k_{3}, \nu_{1},a_{1},\nu_{2},a_{2}, \tau_{1}, \tau_{2}, \theta, \phi)=w_{dd}(\lambda_{\nu_{2}}^{a_{2}}) (2\lambda_{\nu_{2}}^{a_{2}}gg_{xy}  - gg_{xxxy})\, .
\ee
When $\nu_{2}=3d_{yy}$,  
\be
B(k_{3}, \nu_{1},a_{1},\nu_{2},a_{2}, \tau_{1}, \tau_{2}, \theta, \phi)=w_{dd}(\lambda_{\nu_{2}}^{a_{2}}) (2 \lambda_{\nu_{2}}^{a_{2}}gg_{xy}  - gg_{xyyy}) \, .
\ee
When $\nu_{2}=3d_{zz}$,  
\be
B(k_{3}, \nu_{1},a_{1},\nu_{2},a_{2}, \tau_{1}, \tau_{2}, \theta, \phi)=w_{dd}(\lambda_{\nu_{2}}^{a_{2}}) (2\lambda_{\nu_{2}}^{a_{2}}gg_{xy} - gg_{xyzz}) \, .
\ee
When $\nu_{2}=3d_{xy}$, 
\be
B(k_{3}, \nu_{1},a_{1},\nu_{2},a_{2}, \tau_{1}, \tau_{2}, \theta, \phi)=-w_{d}(\lambda_{\nu_{2}}^{a_{2}}) gg_{xxyy} \,.
\ee 
When $\nu_{2}=3d_{xz}$, 
\be
B(k_{3}, \nu_{1},a_{1},\nu_{2},a_{2}, \tau_{1}, \tau_{2}, \theta, \phi)=-w_{d}(\lambda_{\nu_{2}}^{a_{2}}) gg_{xxyz} \,.
\ee
When $\nu_{2}=3d_{yz}$, 
\be
B(k_{3}, \nu_{1},a_{1},\nu_{2},a_{2}, \tau_{1}, \tau_{2}, \theta, \phi)=-w_{d}(\lambda_{\nu_{2}}^{a_{2}}) gg_{xyyz} \,.
\ee

\subsection{For {$\nu_{1}=3d_{xz}$}} 
For $\nu_{1}=3d_{xz}$, 
\be
A(\nu_{1},a_{1})= -w_{d}(\lambda_{\nu_{1}}^{a_{1}}) \,.
\ee

When $\nu_{2}=1s6, 2s3, 2s1$, 
\be
B(k_{3}, \nu_{1},a_{1},\nu_{2},a_{2}, \tau_{1}, \tau_{2}, \theta, \phi)= w_{s}(\lambda_{\nu_{2}}^{a_{2}}) gg_{xz} \,.
\ee
When $\nu_{2}=2p_{x}3, 2p_{x}1$,
\be
B(k_{3}, \nu_{1},a_{1},\nu_{2},a_{2}, \tau_{1}, \tau_{2}, \theta, \phi)=-i w_{p}(\lambda_{\nu_{2}}^{a_{2}}) gg_{xxz}  \,.
\ee 
When $\nu_{2}=2p_{y}3, 2p_{y}1$,
\be
B(k_{3}, \nu_{1},a_{1},\nu_{2},a_{2}, \tau_{1}, \tau_{2}, \theta, \phi)=-i w_{p}(\lambda_{\nu_{2}}^{a_{2}}) gg_{xyz} \,.
\ee 
When $\nu_{2}=2p_{z}3, 2p_{z}1$,
\be
B(k_{3}, \nu_{1},a_{1},\nu_{2},a_{2}, \tau_{1}, \tau_{2}, \theta, \phi)=-i w_{p}(\lambda_{\nu_{2}}^{a_{2}}) gg_{xzz} \,.
\ee 
When $\nu_{2}=3d_{xx}$,  
\be
B(k_{3}, \nu_{1},a_{1},\nu_{2},a_{2}, \tau_{1}, \tau_{2}, \theta, \phi)=w_{dd}(\lambda_{\nu_{2}}^{a_{2}}) (2\lambda_{\nu_{2}}^{a_{2}}gg_{xz}  - gg_{xxxz})\, .
\ee
When $\nu_{2}=3d_{yy}$,  
\be
B(k_{3}, \nu_{1},a_{1},\nu_{2},a_{2}, \tau_{1}, \tau_{2}, \theta, \phi)=w_{dd}(\lambda_{\nu_{2}}^{a_{2}}) (2 \lambda_{\nu_{2}}^{a_{2}}gg_{xz}  - gg_{xyyz}) \, .
\ee
When $\nu_{2}=3d_{zz}$,  
\be
B(k_{3}, \nu_{1},a_{1},\nu_{2},a_{2}, \tau_{1}, \tau_{2}, \theta, \phi)=w_{dd}(\lambda_{\nu_{2}}^{a_{2}}) (2\lambda_{\nu_{2}}^{a_{2}}gg_{xz} - gg_{xzzz}) \, .
\ee
When $\nu_{2}=3d_{xy}$, 
\be
B(k_{3}, \nu_{1},a_{1},\nu_{2},a_{2}, \tau_{1}, \tau_{2}, \theta, \phi)=-w_{d}(\lambda_{\nu_{2}}^{a_{2}}) gg_{xxyz} \,.
\ee 
When $\nu_{2}=3d_{xz}$, 
\be
B(k_{3}, \nu_{1},a_{1},\nu_{2},a_{2}, \tau_{1}, \tau_{2}, \theta, \phi)=-w_{d}(\lambda_{\nu_{2}}^{a_{2}}) gg_{xxzz} \,.
\ee
When $\nu_{2}=3d_{yz}$, 
\be
B(k_{3}, \nu_{1},a_{1},\nu_{2},a_{2}, \tau_{1}, \tau_{2}, \theta, \phi)=-w_{d}(\lambda_{\nu_{2}}^{a_{2}}) gg_{xyzz} \,.
\ee

\subsection{For {$\nu_{1}=3d_{yz}$}} 
For $\nu_{1}=3d_{yz}$, 
\be
A(\nu_{1},a_{1})= -w_{d}(\lambda_{\nu_{1}}^{a_{1}}) \,.
\ee

When $\nu_{2}=1s6, 2s3, 2s1$, 
\be
B(k_{3}, \nu_{1},a_{1},\nu_{2},a_{2}, \tau_{1}, \tau_{2}, \theta, \phi)= w_{s}(\lambda_{\nu_{2}}^{a_{2}}) gg_{yz} \,.
\ee
When $\nu_{2}=2p_{x}3, 2p_{x}1$,
\be
B(k_{3}, \nu_{1},a_{1},\nu_{2},a_{2}, \tau_{1}, \tau_{2}, \theta, \phi)=-i w_{p}(\lambda_{\nu_{2}}^{a_{2}}) gg_{xyz}  \,.
\ee 
When $\nu_{2}=2p_{y}3, 2p_{y}1$,
\be
B(k_{3}, \nu_{1},a_{1},\nu_{2},a_{2}, \tau_{1}, \tau_{2}, \theta, \phi)=-i w_{p}(\lambda_{\nu_{2}}^{a_{2}}) gg_{yyz} \,.
\ee 
When $\nu_{2}=2p_{z}3, 2p_{z}1$,
\be
B(k_{3}, \nu_{1},a_{1},\nu_{2},a_{2}, \tau_{1}, \tau_{2}, \theta, \phi)=-i w_{p}(\lambda_{\nu_{2}}^{a_{2}}) gg_{yzz} \,.
\ee 
When $\nu_{2}=3d_{xx}$,  
\be
B(k_{3}, \nu_{1},a_{1},\nu_{2},a_{2}, \tau_{1}, \tau_{2}, \theta, \phi)=w_{dd}(\lambda_{\nu_{2}}^{a_{2}}) (2\lambda_{\nu_{2}}^{a_{2}}gg_{yz}  - gg_{xxyz})\, .
\ee
When $\nu_{2}=3d_{yy}$,  
\be
B(k_{3}, \nu_{1},a_{1},\nu_{2},a_{2}, \tau_{1}, \tau_{2}, \theta, \phi)=w_{dd}(\lambda_{\nu_{2}}^{a_{2}}) (2 \lambda_{\nu_{2}}^{a_{2}}gg_{yz}  - gg_{yyyz}) \, .
\ee
When $\nu_{2}=3d_{zz}$,  
\be
B(k_{3}, \nu_{1},a_{1},\nu_{2},a_{2}, \tau_{1}, \tau_{2}, \theta, \phi)=w_{dd}(\lambda_{\nu_{2}}^{a_{2}}) (2\lambda_{\nu_{2}}^{a_{2}}gg_{yz} - gg_{yzzz}) \, .
\ee
When $\nu_{2}=3d_{xy}$, 
\be
B(k_{3}, \nu_{1},a_{1},\nu_{2},a_{2}, \tau_{1}, \tau_{2}, \theta, \phi)=-w_{d}(\lambda_{\nu_{2}}^{a_{2}}) gg_{xyyz} \,.
\ee 
When $\nu_{2}=3d_{xz}$, 
\be
B(k_{3}, \nu_{1},a_{1},\nu_{2},a_{2}, \tau_{1}, \tau_{2}, \theta, \phi)=-w_{d}(\lambda_{\nu_{2}}^{a_{2}}) gg_{xyzz} \,.
\ee
When $\nu_{2}=3d_{yz}$, 
\be
B(k_{3}, \nu_{1},a_{1},\nu_{2},a_{2}, \tau_{1}, \tau_{2}, \theta, \phi)=-w_{d}(\lambda_{\nu_{2}}^{a_{2}}) gg_{yyzz} \,.
\ee

\bibliographystyle{jpsj}
\bibliography{72037}

\begin{thebibliography}{10}

\bibitem{cooper85}
M.~J. Cooper: Reports on Progress in Physics {\bfseries 48} (1985) 415.

\bibitem{cooper}
M.~Cooper, P.~Mijnarends, N.~Shiotani, N.~Sakai, and A.~Bansil: {\em X-Ray
  Compton Scattering} (Oxford University Press, Oxford. U.K., 2004).

\bibitem{sakurai16}
H.~Sakurai, N.~Kawahara, M.~Itou, E.~Tomita, K.~Suzuki, and Y.~Sakurai: Journal
  of Synchrotron Radiation {\bfseries 23} (2016) 617.

\bibitem{sakurai21}
H.~Sakurai, N.~Tsuji, Y.~Zama, K.~Suzuki, K.~Hoshi, D.~Hiramoto, Y.~Sakurai,
  and T.~Furuhata: Crystals {\bfseries 11} (2021) 787.

\bibitem{mizuno26}
Y.~Mizuno, H.~Suzuki, N.~Tsuji, T.~Oguchi, G.~Matsubara, H.~Yamase, and
  Y.~Sakurai: Journal of Synchrotron Radiation {\bfseries 33} (2026) 788.

\bibitem{yamase26a}
{H. Yamase, Y. Mizuno, H. Suzuki, N. Tsuji, T. Oguchi, G. Matsubara, and Y.
  Sakurai}: to be published in Communications Engineering  (2026).

\bibitem{manninen77}
S.~Manninen, T.~Paakkari, and V.~Halonen: Chemical Physics Letters {\bfseries
  46} (1977) 62.

\bibitem{hakala06}
M.~Hakala, K.~Nyg{\aa}rd, S.~Manninen, S.~Huotari, T.~Buslaps, A.~Nilsson,
  L.~G.~M. Pettersson, and K.~H\"{a}m\"{a}l\"{a}inen: The Journal of Chemical
  Physics {\bfseries 125} (2006) 084504.

\bibitem{ono16}
T.~Ono, M.~Watanabe, Y.~Sato, H.~Inomata, K.~Nakahara, M.~Itou, and Y.~Sakurai:
  Review of Scientific Instruments {\bfseries 87} (2016) 083106.

\bibitem{saenz97}
A.~Saenz, T.~Asthalter, and W.~Weyrich: International Journal of Quantum
  Chemistry {\bfseries 65} (1997) 213.

\bibitem{lehtola11}
J.~Lehtola, M.~Hakala, J.~Vaara, and K.~H\"{a}m\"{a}l\"{a}inen: Physical
  Chemistry Chemical Physics {\bfseries 13} (2011) 5630.

\bibitem{barca20}
G.~M.~J. Barca, C.~Bertoni, L.~Carrington, D.~Datta, N.~De~Silva, J.~E.
  Deustua, D.~G. Fedorov, J.~R. Gour, A.~O. Gunina, E.~Guidez, T.~Harville,
  S.~Irle, J.~Ivanic, K.~Kowalski, S.~S. Leang, H.~Li, W.~Li, J.~J. Lutz,
  I.~Magoulas, J.~Mato, V.~Mironov, H.~Nakata, B.~Q. Pham, P.~Piecuch,
  D.~Poole, S.~R. Pruitt, A.~P. Rendell, L.~B. Roskop, K.~Ruedenberg,
  T.~Sattasathuchana, M.~W. Schmidt, J.~Shen, L.~Slipchenko, M.~Sosonkina,
  V.~Sundriyal, A.~Tiwari, J.~L. Galvez~Vallejo, B.~Westheimer, M.~Wloch,
  P.~Xu, F.~Zahariev, and M.~S. Gordon: The Journal of Chemical Physics
  {\bfseries 152} (2020) 154102.

\end{thebibliography}


\end{document}